\documentclass[journal,twoside]{IEEEtran}

\usepackage{cite} %
\usepackage{graphicx} %
\usepackage{amsmath,amsfonts,amsbsy,amssymb,amsthm}
\usepackage[caption=false,font=scriptsize]{subfig} %
\usepackage{url} %
\usepackage{multirow}
\usepackage{stfloats} %
\usepackage{booktabs}

\usepackage[usenames]{color}
\usepackage{arydshln}
\def\dashvertical{;{2pt/3pt}}
\def\dashhorizontal{\hdashline[2pt/3pt]}

\DeclareMathAlphabet{\mathpzc}{OT1}{pzc}{m}{it}

\DeclareSymbolFont{bbold}{U}{bbold}{m}{n}
\DeclareSymbolFontAlphabet{\mathbbold}{bbold}
\newcommand{\II}[1]{{\mathbbold{1}\left[#1\right]}}

\usepackage[T1]{fontenc}
\usepackage{textcomp}
\usepackage[scaled=0.78]{beramono}

\renewcommand{\binom}[2]{\left(\genfrac{}{}{0pt}{}{#1}{#2}\right)}

\newcommand{\BB}[2]{{\mathcal{B}\left(#1,#2\right)}}
\newcommand{\ceil}[1]{{\left\lceil{#1}\right\rceil}}
\newcommand{\ee}{{\mathrm{e}}}
\newcommand{\EE}[1]{{\mathbb{E}\left[#1\right]}}

\newcommand{\floor}[1]{{\left\lfloor{#1}\right\rfloor}}
\newcommand{\m}[1]{{\mbox{#1}}}
\newcommand{\nset}{{\m{$\left\{1,\ldots,n\right\}$}}}
\newcommand{\PP}[1]{{\mathbb{P}\left[#1\right]}}
\newcommand{\PPP}{{\mathbf{\Pi}}}
\newcommand{\Ps}{{P_{\text{S}}}}
\newcommand{\Psm}[1]{{\m{$\Ps\left(p,T,m{=}#1\right)$}}}

\newcommand{\RRR}{{\mathpzc{R}}}
\newcommand{\rr}{{\mathbf{r}}}
\newcommand{\uu}{{\mathbf{u}}}
\newcommand{\vv}{{\mathbf{v}}}

\newcommand{\xxs}{{\bar{\mathbf{x}}}}
\newcommand{\xxsm}[1]{{\m{$\xxs\left(n,T,m{=}#1\right)$}}}
\newcommand{\xxr}{{\m{$\left(\frac{1}{r},\ldots,\frac{1}{r}\right)$}}}
\newcommand{\xxn}{{\m{$\left(x_1,\ldots,x_n\right)$}}}
\newcommand{\ZZ}{{\mathbb{Z}}}

\newcommand{\startcompact}[1]{\par\vspace{-0.75em}\begin{#1}\allowdisplaybreaks\ignorespaces}
\newcommand{\stopcompact}[1]{\end{#1}\ignorespaces}

\allowdisplaybreaks

\theoremstyle{definition}
\newtheorem*{thm:problem}{Problem}
\newtheorem*{thm:definition}{Definition}
\newtheorem*{thm:conjecture}{Conjecture}

\theoremstyle{plain}
\newtheorem{thm:claim}{Claim}
\newtheorem{thm:proposition}{Proposition}
\newtheorem{thm:lemma}{Lemma}
\newtheorem{thm:corollary}{Corollary}
\newtheorem{thm:theorem}{Theorem}

\newcounter{tempeqcounter}

\begin{document}

\title{Distributed Storage Allocations}

\author{%
\m{Derek Leong,}
\m{Alexandros G. Dimakis,}
\m{and Tracey Ho}
\thanks{%
\hrule width 0.33\columnwidth \vskip5pt
This paper is an extended version of \cite{dl:leong12allocations}, which is available online at {\texttt{\protect\url{http://dx.doi.org/10.1109/TIT.2012.2191135}}}.
The material in this paper was presented in part at the
\m{NetCod 2009 \cite{dl:leong09distributed}},
\m{ICC 2010 \cite{dl:leong10distributed}}, and
\m{GLOBECOM 2010 \cite{dl:leong10symmetric}} conferences.}%
\thanks{%
The work of D.~Leong and T.~Ho was supported in part by
Subcontract 069153 issued by BAE Systems National Security Solutions, Inc.~and supported by the Defense Advanced Research Projects Agency (DARPA) and the Space and Naval Warfare System Center (SPAWARSYSCEN), San Diego, under Contract N66001-08-C-2013,
by the Air Force Office of Scientific Research under Grant FA9550-10-1-0166,
and by the Lee Center for Advanced Networking at the California Institute of Technology (Caltech).
The work of D.~Leong was supported in part by A*STAR, Singapore.
The work of A.~G.~Dimakis was supported in part by the Center for the Mathematics of Information at Caltech,
and by the National Science Foundation under NSF Grant 1055099.}%
\thanks{%
D.~Leong and T.~Ho are with the
Department of Electrical \m{Engineering},
California Institute of Technology,
Pasadena, California 91125, USA
\m{(email: \texttt{derekleong@caltech.edu}, \texttt{tho@caltech.edu}).}}%
\thanks{%
A.~G.~Dimakis is with the
Department of Electrical Engineering,
University of Southern California,
Los Angeles, California 90089, USA
\m{(email: \texttt{dimakis@usc.edu}).}}%
} %

\maketitle

\begin{abstract}
We examine the problem of allocating a given total storage budget in a distributed storage system for maximum reliability.
A source has a single data object that is to be coded and stored over a set of storage nodes;
it is allowed to store any amount of coded data in each node, as long as the total amount of storage used does not exceed the given budget.
A data collector subsequently attempts to recover the original data object by accessing only the data stored in a random subset of the nodes.
By using an appropriate code, successful recovery can be achieved whenever the total amount of data accessed is at least the size of the original data object.
The goal is to find an optimal storage allocation that maximizes the probability of successful recovery.
This optimization problem is challenging in general because of its combinatorial nature, despite its simple formulation.
We study several variations of the problem, assuming different allocation models and access models.
The optimal allocation and the optimal \emph{symmetric} allocation (in which all nonempty nodes store the same amount of data) are determined for a variety of cases.
Our results indicate that the optimal allocations often have nonintuitive structure and are difficult to specify.
We also show that depending on the circumstances, coding may or may not be beneficial for reliable storage.
\end{abstract}


\IEEEpeerreviewmaketitle

\section{Introduction}

\IEEEPARstart{C}{onsider} a distributed storage system comprising $n$ storage nodes.
A source has a single data object of normalized unit size that is to be coded and stored in a distributed manner over these nodes, subject to a given total storage budget $T$.
Let $x_i$ be the amount of coded data stored in node $i\in$ $\nset$.
Any amount of data may be stored in each node, as long as the total amount of storage used over all nodes is at most the given budget $T$, i.e.,
\[
\sum_{i=1}^{n} x_i \leq T.
\]
This is a realistic constraint if there is limited transmission bandwidth or storage space, or if it is too costly to mirror the data object in its entirety in every node.
At some time after the creation of this coded storage, a data collector attempts to recover the original data object by accessing only the data stored in a \emph{random} subset $\rr$ of the nodes, where the probability distribution of \m{$\rr\subseteq\nset$} is specified by an assumed access model or failure model (nodes or links may fail probabilistically, for example).
\begin{figure}
\centering
\includegraphics[width=0.27\textwidth]{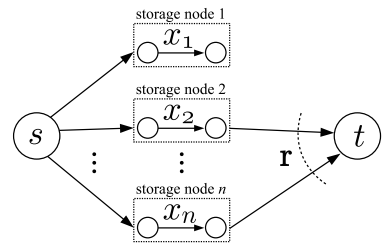}
\caption{
Information flows in a distributed storage system.
The source $s$ has a single data object of normalized unit size that is to be coded and stored over $n$ storage nodes.
Subsequently, a data collector $t$ attempts to recover the original data object by accessing only the data stored in a random subset $\rr$ of the nodes.}
\label{fig:NetworkRandomSubset}
\end{figure}%
Fig.~\ref{fig:NetworkRandomSubset} depicts such a distributed storage system.

The \emph{reliability} of this system, which we define to be the probability of successful recovery (or recovery probability in short), depends on both the storage allocation and the coding scheme.
For maximum reliability, we would therefore need to find
\begin{enumerate}
\item
an optimal allocation of the given budget $T$ over the nodes, specified by the values of $x_1,\ldots,x_n$, and
\item
an optimal coding scheme
\end{enumerate}
that jointly maximize the probability of successful recovery.
It turns out that these two problems can be decoupled by using a good coding scheme, specifically one that enables successful recovery whenever the total amount of data accessed by the data collector is at least the size of the original data object.
This can be seen by considering the information flows for a network in which the source is multicasting the data object to a set of potential data collectors \cite{dl:dimakis10network,dl:jiang06network}:
successful recovery can be achieved by a data collector if and only if its corresponding max-flow or min-cut from the source is at least the size of the original data object.
Random linear coding over a sufficiently large field would allow successful recovery with high probability when this condition is satisfied \cite{dl:ho06random,dl:fragouli06network}.
Alternatively, a suitable maximum distance separable (MDS) code for the given budget and data object size would allow successful recovery with certainty when this condition is satisfied.

Therefore, assuming the use of an appropriate code, the probability of successful recovery for an allocation $\xxn$ can be written as
\[
\PP{\text{successful recovery}} = \PP{\sum_{i \in \rr} x_i \geq 1}.
\]
Our goal is to find an optimal allocation that maximizes this recovery probability, subject to the given budget constraint.

Although we have assumed coded storage at the outset, coding may ultimately be unnecessary for certain allocations.
For example, if the budget is spread minimally such that each nonempty node stores the data object in its entirety (i.e., \m{$x_i\geq 1$} for all \m{$i\in S$}, and \m{$x_i=0$} for all \m{$i\notin S$}, where $S$ is some subset of $\nset$), then uncoded replication would suffice since the data object can be recovered by accessing any \emph{one} nonempty node;
the data collector would not need to combine data accessed from different nodes in order to recover the data object.
Thus, by solving for the optimal allocation, we will also be able to determine whether coding is beneficial for reliable storage.

We note that even though no explicit upper bound is imposed on the amount of data that can be stored in each node, it is never necessary to set \m{$x_i>1$} because \m{$x_i=1$} already allows the data object to be stored in its entirety in that node.
The absence of a tighter per-node storage constraint \m{$x_i\leq c_i<1$} is reasonable for storage systems that handle a large number of data objects:
we would expect the storage capacity of each node to be much larger than the size of a single data object,
making it possible for a node to accommodate some of the data objects in their entirety.
As such, it would be appropriate to apply a storage constraint for each data object via the budget $T$, without a separate \textit{a priori} constraint for $x_i$.
Furthermore, the simplifying assumption of $x_i$ being a continuous variable is a reasonable one for large data objects:
a large data object size would facilitate the creation of coded data packets with sizes (closely) matching that of a desired allocation.
Incidentally, the overhead associated with random linear coding or an MDS code, which is ignored in our model, becomes proportionately negligible when the amount of coded data is large.

In spite of the simple formulation, this optimization problem poses significant challenges because of its combinatorial nature and the large space of feasible allocations.
Different variations of this problem can be formulated by assuming different allocation models and access models;
in this paper, we will examine three such variations that are motivated by practical storage problems in content delivery networks, delay tolerant networks, and wireless sensor networks.

\subsection{Independent Probabilistic Access to Each Node}

In the first problem formulation, we assume that the data collector accesses each of the $n$ nodes independently with constant probability $p$;
in other words, each node $i$ appears in subset $\rr$ independently with probability $p$.
The resulting problem can be interpreted as that of maximizing the reliability of data storage in a system comprising $n$ storage devices where each device fails independently with probability \m{$1-p$}.
It is not hard to show that determining the recovery probability of a \emph{given} allocation is computationally difficult (specifically, \#P-hard).
The intuitive approach of spreading the budget maximally over all nodes, i.e., setting \m{$x_i=\frac{T}{n}$} for all $i$, turns out to be not necessarily optimal;
in fact, the optimal allocation may not even be symmetric
(we say that an allocation is \emph{symmetric} when all nonzero $x_i$ are equal).
The following counterexample from \cite{dl:karppersonal} demonstrates that symmetric allocations can be suboptimal:
for \m{$(n,p,T)$} $=$ \m{$\left(5,\frac{2}{3},\frac{7}{3}\right)$},
the nonsymmetric allocation
\[
\textstyle
\left(\frac{2}{3},\frac{2}{3},\frac{1}{3},\frac{1}{3},\frac{1}{3}\right),
\]
which achieves a recovery probability of $0.90535$, performs strictly better than any symmetric allocation;
the maximum recovery probability among symmetric allocations is $0.88889$, which is achieved by both
\[
\textstyle
\left(\frac{7}{6},\frac{7}{6},0,0,0\right)
\text{ and }
\left(\frac{7}{12},\frac{7}{12},\frac{7}{12},\frac{7}{12},0\right).
\]
Evidently, the simple strategy of ``spreading eggs evenly over more baskets'' may not always improve the reliability of an allocation.

\textit{\textbf{Our Contribution}}:
We show that the intuitive symmetric allocation that spreads the budget maximally over all nodes is indeed \emph{asymptotically} optimal in a regime of interest.
Specifically, we derive an upper bound for the suboptimality of this allocation, and show that the performance gap vanishes asymptotically as the total number of storage nodes $n$ grows, when \m{$p>\frac{1}{T}$}.
This is a regime of interest because a high recovery probability is possible when \m{$p>\frac{1}{T}$} $\Longleftrightarrow$ \m{$pT>1$}:
The expected total amount of data accessed by the data collector is given by
\begin{align}
\EE{\displaystyle\sum_{i=1}^{n} x_i Y_i}
=\sum_{i=1}^{n} x_i \EE{Y_i}
=p \sum_{i=1}^{n} x_i
\leq pT,
\label{eq:IndeptProbAccessExpTotal}
\end{align}
where \m{$Y_i$'s} are independent \m{$\text{Bernoulli}(p)$} random variables.
Therefore, the data collector would be able to access a sufficient amount of data \emph{in expectation} for successful recovery if \m{$pT>1$}.

We also show that the symmetric allocation that spreads the budget minimally is optimal when $p$ is sufficiently small.
In such an allocation, the data object is stored in its entirety in each nonempty node, making coding unnecessary.
Additionally, we explicitly find the optimal \emph{symmetric} allocation for a wide range of parameter values of $p$ and $T$.

\textit{\textbf{Related Work}}:
This problem was introduced to us through a discussion at UC Berkeley \cite{dl:karppersonal}.
We have since learned that variations of the problem have also been studied in several different fields.

In reliability engineering, the weighted-$k$-out-of-$n$ system \cite{dl:wu94reliability} comprises $n$ components, each having a positive integer weight $w_i$ and surviving independently with probability $p_i$;
the system is in a good state if and only if the total weight of its surviving components is at least a specified threshold $k$.
Related work on this system and its extensions has focused on the efficient computation of the reliability of a given weight allocation (see, e.g., \cite{dl:chen05reliability}).

In peer-to-peer networking, the allocation problem deals with the recovery of a data object from peers that are available only probabilistically.
Lin \textit{et al.}~\cite{dl:lin04erasure} compared the performance of uncoded replication vs.~coded storage, restricted to symmetric allocations, for the case where the budget is an integer.

In wireless communications, the allocation problem is studied in the context of multipath routing, in which coded data is transmitted along different paths in an unreliable network, exploiting path diversity to improve the reliability of end-to-end communications.
Tsirigos and Haas \cite{dl:tsirigos04multipath1,dl:tsirigos04multipath2} examined the performance of symmetric allocations and noted the existence of a phase transition in the optimal symmetric allocation;
approximation methods were also proposed by the authors to tackle the optimization problem, especially for the case where path failures occur with nonuniform probabilities and may be correlated.
Jain \textit{et al.}~\cite{dl:jain05using} evaluated the performance of symmetric allocations experimentally in a delay tolerant network setting, and presented an alternative formulation using Gaussian distributions to model partial access to nodes.

Our work generalizes these previous efforts by considering nonsymmetric allocations and noninteger budgets.
We also correct some inaccurate claims about the optimal symmetric allocation in \cite{dl:jain05using} and its associated technical report.

\subsection{Access to a Random Fixed-Size Subset of Nodes}

In the second problem formulation, we assume that the data collector accesses an \m{$r$-subset} of the $n$ nodes selected uniformly at random from the collection of all $\binom{n}{r}$ possible \m{$r$-subsets}, where $r$ is a given constant.
The resulting problem can be interpreted as that of maximizing the recovery probability in a networked storage system of $n$ nodes where the end user is able or allowed to contact up to $r$ nodes randomly.
We can treat this access model as an approximation to the preceding independent probabilistic access model by picking \m{$r\approx np$}.
Finding the optimal allocation in this case is still challenging.
As in the first problem formulation, it is not hard to show that determining the recovery probability of a \emph{given} allocation is computationally difficult (specifically, \#P-complete).

The problem appears nontrivial even if we restrict the optimization to only \emph{symmetric} allocations.
Numerically, we observe that given $n$ and $r$, either a minimal or a maximal spreading of the budget is optimal among symmetric allocations for most, if not all, choices of $T$.
One example of an exception is \m{$(n,r,T)$} $=$ \m{$\left(14,5,\frac{8}{3}\right)$} for which it is optimal to have $8$ nonempty nodes in the symmetric allocation, instead of the extremes $2$ or $13$;
another example is \m{$(n,r,T)$} $=$ \m{$\left(16,4,\frac{7}{2}\right)$} for which it is optimal to have $7$ nonempty nodes in the symmetric allocation, instead of the extremes $3$ or $14$.
Furthermore, the number of nonempty nodes in the optimal symmetric allocation is not necessarily a nondecreasing function of the budget $T$;
for instance, given
\m{$(n,r)$} $=$ \m{$(20,4)$},
it is optimal to have
\m{$(4,18,14,19,20)$} nonempty nodes in the symmetric allocation for
\m{$T=(4.25,4.5,4.67,4.75,5)$}, respectively.

\textit{\textbf{Our Contribution}}:
We show that the allocation $\xxr$ is optimal in the \emph{high recovery probability regime}.
Specifically, we demonstrate that this allocation, which has a recovery probability of exactly $1$, minimizes the budget $T$ necessary for achieving any recovery probability exceeding a specified threshold \m{$1-\epsilon$}.
Although $\epsilon$ depends on $n$ and $r$ in a complicated way, we can conclude that for any $r$, this allocation is optimal if the recovery probability is to exceed \m{$1-\frac{1}{n}$}.

We also make the following conjecture about the optimal allocation, based on our numerical observations:

\begin{thm:conjecture}
A \emph{symmetric} optimal allocation always exists for any $n$, $r$, and $T$.
\end{thm:conjecture}

\textit{\textbf{Related Work}}:
Sardari \textit{et al.}~\cite{dl:sardari10memory} presented a method of \emph{approximating} an optimal solution to this problem by considering a data collector that accesses $r$ random nodes with replacement.
More recently, Alon \textit{et al.}~\cite{dl:alon11matchings} showed that this problem is related to an old conjecture by Erd\H{o}s on the maximum number of edges in a uniform hypergraph \cite{dl:erdos65tuples}.

\subsection{Probabilistic Symmetric Allocations}

In the third problem formulation, we assume a \emph{probabilistic} allocation model in which the source selects a random allocation from a distribution of allocations, with the constraint that the \emph{expected} total amount of storage used in an allocation is at most the given budget $T$.
We specifically consider the case where each of the $n$ nodes is selected by the source independently with constant probability
\m{$\min\!\left(\frac{\ell T}{n},1\right)$}
to store a constant $\frac{1}{\ell}$ amount of data, thus creating a probabilistic \emph{symmetric} allocation of the budget.
The data collector subsequently accesses an \m{$r$-subset} of the $n$ nodes selected uniformly at random from the collection of all $\binom{n}{r}$ possible \m{$r$-subsets}, where $r$ is a given constant.
The goal is to find an optimal allocation, specified by the value of parameter $\ell$, that maximizes the recovery probability.
This model was conceived as a simplification of the preceding fixed-size subset access model which assumes a deterministic allocation of the budget.

\textit{\textbf{Our Contribution}}:
We show that the choice of \m{$\ell=r$}, which corresponds to a maximal spreading of the budget, is optimal when the given budget $T$ is sufficiently large, or equivalently, when a sufficiently high recovery probability (specifically, $\frac{3}{4}$ or higher) is achievable.
We believe this is a reasonable operating regime for applications that require good reliability.

\subsection{Other Related Work}

Apart from the work done on the preceding problems, a variety of storage allocation problems have also been studied in a \emph{nonprobabilistic} setting.
For instance, the objective adopted in \cite{dl:naor95optimal} and \cite{dl:jiang05network} is to minimize the total storage budget required to satisfy a given set of deterministic recovery requirements in a network.
Incidentally, the use of network coding makes it easier to deal with the total cost of content delivery, which covers the initial dissemination, storage, and eventual fetching of data objects;
this cost-minimization problem is considered in \cite{dl:jiang06network} and \cite{dl:leong09optimal}, subject to various deterministic constraints involving, for example, load balancing or fetching distance.

We note that in most of the literature involving reliable distributed storage, either the data object is assumed to be replicated in its entirety (see, e.g., \cite{dl:spyropoulos05spray}), or, if coding is used, every node is assumed to store the same amount of coded data (see, e.g., \cite{dl:acedanski05how,dl:dimakis05ubiquitous,dl:kamra06growth,dl:lin07data,dl:aly08fountain}).
Allocations of a storage budget with nodes possibly storing different amounts of data are not usually considered.

In the following three sections, we define each problem formally and state our main results.
Proofs of theorems are deferred to the appendix.
Table~\ref{tbl:Notation} summarizes the notation used throughout this paper.

\begin{table}
\caption{%
Notation}
\label{tbl:Notation}
\centering\small
\begin{tabular}{c@{\hspace{0.5em}}l}
\toprule
Symbol & Definition
\\ \midrule
$n$ & total number of storage nodes, $n\geq 2$
\\
$x_i$ & amount of data stored in storage node $i$,
\\
~ & \hspace{0.5em} $x_i \geq 0$, where $i\in$ $\nset$
\\
$T$ & total storage budget, $1 \leq T \leq n$
\\
$\rr$ & subset of nodes accessed, $\rr\subseteq$ $\nset$
\\
$p$ & access probability (Section~\ref{sec:IndeptProbAccess}), $0 < p < 1$
\\
$r$ & number of nodes accessed (Section~\ref{sec:RandFixedSizeSubset}), $1\leq r\leq n$
\\
$\frac{1}{\ell}$ & amount of data stored in each nonempty node
\\
~ & \hspace{0.5em} (Section~\ref{sec:ProbAlloc}), $\ell >0$
\\
$\BB{n}{p}$ & binomial random variable with $n$ trials and
\\
~ & \hspace{0.5em} success probability $p$
\\
$\II{G}$ & indicator function; $\II{G}=1$ if statement $G$ is true,
\\
~ & \hspace{0.5em} and $0$ otherwise
\\
$\ZZ^+_0$ & the set of nonnegative integers, i.e., $\ZZ^+\cup\{0\}$
\\ \bottomrule
\end{tabular}
\end{table}

\section{Independent Probabilistic Access to Each Node}
\label{sec:IndeptProbAccess}

In the first variation of the storage allocation problem, we consider a data collector that accesses each of the $n$ nodes independently with probability $p$;
successful recovery occurs if and only if the total amount of data stored in the accessed nodes is at least $1$.
We seek an optimal allocation $\xxn$ of the budget $T$ that maximizes the probability of successful recovery, for a given choice of $n$, $p$, and $T$.
This optimization problem can be expressed as follows:
\begin{align}
& \hspace{-0.3em} \PPP_1(n,p,T): \notag
\\*
& \hspace{0.4em} \underset{x_1,\ldots,x_n}{\text{maximize}} \sum_{\rr\subseteq\{1,\ldots,n\}} p^{|\rr|} (1-p)^{n-|\rr|}
\cdot \II{\sum_{i\in\rr} x_i \geq 1} \label{eq:P1ObjFunc}
\\*
& \hspace{0.4em} \text{subject to} \notag
\\*
& \hspace{4.65em} \sum_{i=1}^{n} x_i \hspace{0em} \leq \hspace{0.5em} T \label{eq:P1Constraint1}
\\*
& \hspace{5.9em} x_i \hspace{0.5em} \geq \hspace{0.5em} 0 \hspace{2.05em} \forall\;\; i\in\nset. \label{eq:P1Constraint2}
\end{align}
The objective function~\eqref{eq:P1ObjFunc} is just the recovery probability, expressed as the sum of the probabilities corresponding to the subsets $\rr$ that allow successful recovery.
An equivalent expression for \eqref{eq:P1ObjFunc} is
\[
\PP{\sum_{i=1}^{n} x_i\,Y_i \geq 1},
\]
where \m{$Y_i$'s} are independent \m{$\text{Bernoulli}(p)$} random variables.
Inequality~\eqref{eq:P1Constraint1} expresses the budget constraint, and inequality~\eqref{eq:P1Constraint2} ensures that a nonnegative amount of data is stored in each node.
For the trivial budget \m{$T=1$}, the allocation \m{$(1,0,\ldots,0)$} is optimal;
for \m{$T=n$}, the allocation \m{$(1,\ldots,1)$} is optimal.
Incidentally, computing the recovery probability of a \emph{given} allocation turns out to be \#P-hard:
\begin{thm:proposition}
\label{thm:proposition:IndeptProbAccessPHard}
Computing the recovery probability
\[
\sum_{\rr\subseteq\{1,\ldots,n\}} p^{|\rr|} (1-p)^{n-|\rr|}
\cdot \II{\sum_{i\in\rr} x_i \geq 1}
\]
for a given allocation $\xxn$ and choice of $p$ is \#P-hard.
\end{thm:proposition}

\begin{table}
\caption{%
Optimal Allocations for Number of Nodes $n=2,3,4$}
\label{tbl:IndependentProbabilisticAccessCompleteSolutions}
\centering\small
\begin{tabular}{l@{\hspace{1em}}c@{\hspace{0.5em}}l@{\hspace{0.5em}}l}
\toprule
$n$ &
\hspace{-0em} Budget $T$ &
\hspace{-0.8em} $\begin{array}{l}
\text{Optimal}\\\text{allocation}
\end{array}$ &
\hspace{-1em} $\begin{array}{l}
\text{Condition on access probability $p$}\\\text{(if any)}
\end{array}$\hspace{-0.5em}
\\ \midrule
\multirow{1}{*}{$2$} & $1\leq T<2$ & $\left(1,0\right)$ & ~
\\ \midrule
\multirow{5}{*}{$3$} & $1\leq T<\frac{3}{2}$ & $\left(1,0,0\right)$ & ~
\\ \cmidrule{2-4}
& \multirow{2}{*}{$\frac{3}{2}\leq T<2$} & $\left(1,0,0\right)$ & \text{if } $p\leq\frac{1}{2}$
\\
& & $\left(\frac{1}{2},\frac{1}{2},\frac{1}{2}\right)$ & \text{if } $p\geq\frac{1}{2}$
\\ \cmidrule{2-4}
& $2\leq T<3$ & $\left(1,1,0\right)$ & ~
\\ \midrule
\multirow{12}{*}{$4$} & $1\leq T<\frac{4}{3}$ & $\left(1,0,0,0\right)$ & ~
\\ \cmidrule{2-4}
& \multirow{2}{*}{$\frac{4}{3}\leq T<\frac{3}{2}$} & $\left(1,0,0,0\right)$ & \text{if } $p\leq\frac{1+\sqrt{13}}{6}\approx 0.768$
\\
& & $\left(\frac{1}{3},\frac{1}{3},\frac{1}{3},\frac{1}{3}\right)$ & \text{if } $p\geq\frac{1+\sqrt{13}}{6}\approx 0.768$
\\ \cmidrule{2-4}
& \multirow{2}{*}{$\frac{3}{2}\leq T<2$} & $\left(1,0,0,0\right)$ & \text{if } $p\leq\frac{1}{2}$
\\
& & $\left(\frac{1}{2},\frac{1}{2},\frac{1}{2},0\right)$ & \text{if } $p\geq\frac{1}{2}$
\\ \cmidrule{2-4}
& \multirow{2}{*}{$2\leq T<\frac{5}{2}$} & $\left(1,1,0,0\right)$ & \text{if } $p\leq\frac{2}{3}$
\\
& & $\left(\frac{1}{2},\frac{1}{2},\frac{1}{2},\frac{1}{2}\right)$ & \text{if } $p\geq\frac{2}{3}$
\\ \cmidrule{2-4}
& \multirow{2}{*}{$\frac{5}{2}\leq T<3$} & $\left(1,1,0,0\right)$ & \text{if } $p\leq\frac{1}{2}$
\\
& & $\left(1,\frac{1}{2},\frac{1}{2},\frac{1}{2}\right)$ & \text{if } $p\geq\frac{1}{2}$
\\ \cmidrule{2-4}
& $3\leq T<4$ & $\left(1,1,1,0\right)$ & ~
\\ \bottomrule
\end{tabular}
\end{table}
Table~\ref{tbl:IndependentProbabilisticAccessCompleteSolutions} lists the optimal allocations for $n=2,3,4$, covering all parameter values of \m{$p\in(0,1)$} and \m{$T\in[1,n)$}.
These solutions are obtained by minimizing $T$ for each possible value of the objective function~\eqref{eq:P1ObjFunc}.
We observe that
\begin{enumerate}
\item\label{enum:MinSpreadOpt}
for any $T$, the symmetric allocation \m{$(1,\ldots,1,0,\ldots,0)$}, which corresponds to a minimal spreading of the budget (uncoded replication), appears to be optimal when $p$ is sufficiently small, and
\item\label{enum:SymmAlloc}
the optimal \emph{symmetric} allocation appears to perform well despite being suboptimal in some cases, e.g., when \m{$(n,T)=\left(4,\frac{5}{2}\right)$} and \m{$p>\frac{1}{2}$}.
\end{enumerate}
We will proceed to show that observation~\ref{enum:MinSpreadOpt} is indeed true in Section~\ref{sec:IndeptProbAccessMinSpreadOpt};
the opposite approach of spreading the budget maximally over all nodes turns out to be \emph{asymptotically} optimal when $p$ is sufficiently large, as will be demonstrated in Section~\ref{sec:IndeptProbAccessMaxSpreadOpt}.
Motivated by observation~\ref{enum:SymmAlloc}, we examine the optimization problem restricted to symmetric allocations in Section~\ref{sec:IndeptProbAccessOptSymmAlloc}.

For brevity, let $\xxs(n,T,m)$ denote the \emph{symmetric} allocation for $n$ nodes that uses a total storage of $T$ and contains exactly $m\in\{1,2,\ldots,n\}$ nonempty nodes:
\[
\xxs(n,T,m)
\triangleq
\biggl(\,
\underbrace{\frac{T}{m},\ldots,\frac{T}{m}}_{m \text{ entries}},
\underbrace{0,\ldots,0\vphantom{\frac{T}{m}}}_{(n-m) \text{ entries}}
\!\!\biggr).
\]
Since successful recovery for the symmetric allocation $\xxs(n,T,m)$ occurs if and only if at least \m{$\ceil{1\big/\left(\frac{T}{m}\right)}$} $=\ceil{\frac{m}{T}}$ out of the $m$ nonempty nodes are accessed, the corresponding probability of successful recovery can be written as
\[
\Ps(p,T,m) \triangleq \PP{\displaystyle\BB{m}{p}\geq\ceil{\frac{m}{T}}}.
\]

\subsection{Asymptotic Optimality of Maximal Spreading}
\label{sec:IndeptProbAccessMaxSpreadOpt}

The recovery probability of the symmetric allocation \xxsm{n}, which corresponds to a maximal spreading of the budget over all nodes, is given by
\begin{align}
\Ps(p,T,m{=}n) =
\PP{\displaystyle\BB{n}{p}\geq\ceil{\frac{n}{T}}}.
\label{eq:MaxSpreadRecoveryProb}
\end{align}
To establish the optimality of this allocation, we compare \eqref{eq:MaxSpreadRecoveryProb} to an upper bound for the recovery probability of an optimal allocation.
Such a bound can be derived by conditioning on the number of accessed nodes:

\begin{thm:lemma}
\label{thm:lemma:IndeptProbAccessUpperBound}
The probability of successful recovery for an optimal allocation is at most
\begin{align}
\sum_{r=0}^{n} \min\left(\frac{rT}{n},1\right) \; \PP{\BB{n}{p}=r}. \label{eq:RecoveryProbUpperBound}
\end{align}
\end{thm:lemma}

The suboptimality of \xxsm{n} is therefore bounded by the difference between \eqref{eq:MaxSpreadRecoveryProb} and \eqref{eq:RecoveryProbUpperBound}, as given by the following theorem;
when \m{$p>\frac{1}{T}$}, this allocation becomes asymptotically optimal since its suboptimality gap vanishes as $n$ goes to infinity:

\begin{thm:theorem}
\label{thm:theorem:IndeptProbAccessMaxSpreadSubopt}
The gap between the probabilities of successful recovery for an optimal allocation and for the symmetric allocation \xxsm{n}, which corresponds to a maximal spreading of the budget over all nodes, is at most
\[
p\,T\;\PP{\displaystyle\BB{n-1}{p}\leq\ceil{\frac{n}{T}}-2}.
\]
If $p$ and $T$ are fixed such that \m{$p>\frac{1}{T}$}, then this gap approaches zero as $n$ goes to infinity.
\end{thm:theorem}

We note that the regime \m{$p>\frac{1}{T}$} is particularly interesting because it corresponds to the regime of high recovery probability;
the recovery probability would be bounded away from $1$ if \m{$p<\frac{1}{T}$} $\Longleftrightarrow$ \m{$pT<1$} instead.
This follows from the application of Markov's inequality to the random variable $W$ denoting the total amount of data accessed by the data collector, which produces
\[
\PP{W\geq 1}\leq\EE{W}.
\]
Since \m{$\PP{W\geq 1}$} is just the probability of successful recovery, and \m{$\EE{W}\leq pT$} according to \eqref{eq:IndeptProbAccessExpTotal}, we have
\[
\PP{\text{successful recovery}}\leq pT.
\]

\subsection{Optimality of Minimal Spreading (Uncoded Replication)}
\label{sec:IndeptProbAccessMinSpreadOpt}

The recovery probability of the symmetric allocation \xxsm{\floor{T}}, which corresponds to a minimal spreading of the budget, is given by
\begin{align}
\Ps\left(p,T,m{=}\floor{T}\right) =
\PP{\displaystyle\BB{\floor{T}}{p}{\geq}1}
= 1-(1{-}p)^{\floor{T}}.
\label{eq:MinSpreadRecoveryProb}
\end{align}
Recall that coding is unnecessary in such an allocation since the data object is stored in its entirety in each nonempty node.
A sufficient condition for the optimality of this allocation can be found by comparing \eqref{eq:MinSpreadRecoveryProb} to an upper bound for the recovery probabilities of all other allocations.
Our approach is to classify each allocation according to the number of individual nodes that store at least a unit amount of data.
We then find a bound for allocations containing exactly $0$ such nodes, another bound for allocations containing exactly $1$ such node, and so on.
The subsequent comparisons of \eqref{eq:MinSpreadRecoveryProb} to each of these bounds result in the following theorem:

\begin{thm:theorem}
\label{thm:theorem:IndeptProbAccessMinSpreadOpt}
If \m{$1<T<n$} and
\begin{align}
& 1 - (1-p)^{\floor{T}-n} + (n-\ell) \left(\frac{p}{1-p}\right) \notag
\\*
&\;\; + \!\!\sum_{r=2}^{\ceil{\frac{n-\ell}{T-\ell}}-1} \left(1-\frac{T-\ell}{n-\ell}\cdot r\right) \binom{n-\ell}{r} \left(\frac{p}{1-p}\right)^r
\geq 0 \label{eq:IndeptProbAccessMinSpreadOptIneq}
\end{align}
for all \m{$\ell\in\{0,1,\ldots,\floor{T}-1\}$}, then \xxsm{\floor{T}}, which corresponds to a minimal spreading of the budget (uncoded replication), is an optimal allocation.
\end{thm:theorem}

\noindent
The following corollary shows that this allocation is indeed optimal for sufficiently small $p$:

\begin{thm:corollary}
\label{thm:corollary:IndeptProbAccessMinSpreadOpt}
If \m{$1<T<n$} and \m{$p\leq\frac{2}{\left(n-\floor{T}\right)^2}$},
then \xxsm{\floor{T}} is an optimal allocation.
\end{thm:corollary}

\begin{figure}
\centering
\includegraphics[width=0.47\textwidth]{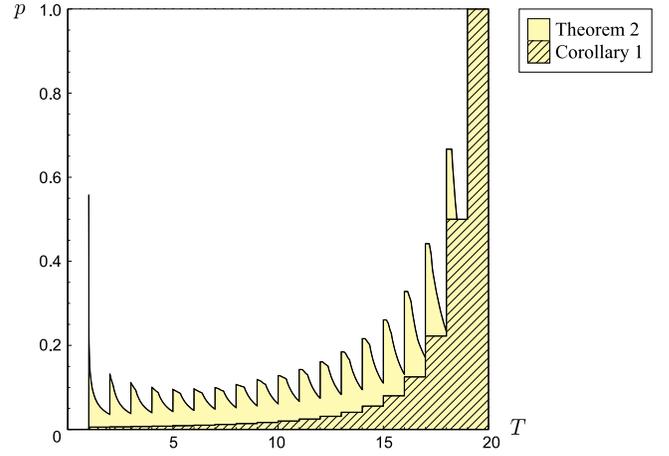}
\caption{Plot of access probability $p$ against budget $T$, showing regions of \m{$(T,p)$} over which the sufficient conditions of the theorems are satisfied, for \m{$n=20$}.
Minimal spreading (uncoded replication) is optimal among all allocations in the colored regions.}
\label{fig:IndeptProbAccessMinSpreadOptRegionPlot}
\end{figure}%
Fig.~\ref{fig:IndeptProbAccessMinSpreadOptRegionPlot} illustrates these results in the form of a region plot for an instance of $n$.

\subsection{Optimal Symmetric Allocation}
\label{sec:IndeptProbAccessOptSymmAlloc}

The optimization problem appears nontrivial even if we were to consider only \emph{symmetric} allocations.
\begin{figure}
\centering
\includegraphics[width=0.47\textwidth]{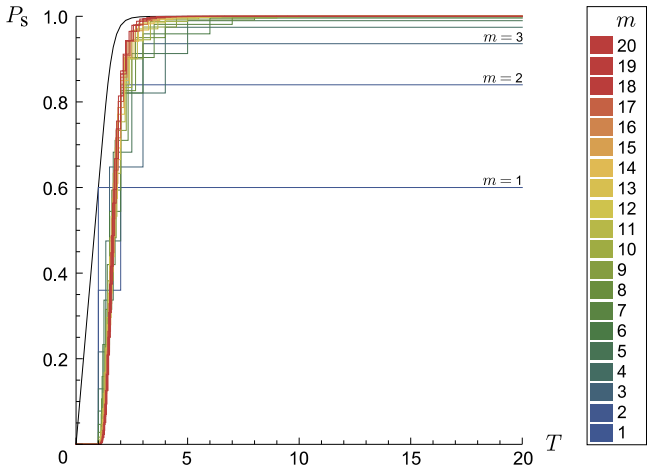}
\caption{Plot of recovery probability $\Ps$ against budget $T$ for each symmetric allocation $\xxs(n,T,m)$, for \m{$(n,p)=\left(20,\frac{3}{5}\right)$}.
Parameter $m$ denotes the number of nonempty nodes in the symmetric allocation.
The black curve gives an upper bound for the recovery probability of an optimal allocation, as derived in Lemma~\ref{thm:lemma:IndeptProbAccessUpperBound}.}
\label{fig:IndeptProbAccessSymmPlot}
\end{figure}%
Fig.~\ref{fig:IndeptProbAccessSymmPlot}, which compares the performance of different symmetric allocations over different budgets for an instance of \m{$(n,p)$}, demonstrates that the value of $m$ corresponding to the optimal symmetric allocation can change drastically as the budget $T$ varies.

Fortunately, we can eliminate many candidates for the optimal value of $m$ by making the following observation:
Recall that the recovery probability of the symmetric allocation $\xxs(n,T,m)$ is given by
\m{$\Ps(p,T,m)$} $\triangleq$
\m{$\PP{\BB{m}{p}\geq\ceil{\frac{m}{T}}}$}.
For fixed $n$, $p$, and $T$, we have
\begin{align*}
\ceil{\frac{m}{T}}
&= k
\hspace{4.5em}
\text{when }
m\in \big((k-1)T,kT\big],
\intertext{for $k = 1,2,\ldots,\floor{\frac{n}{T}}$, and finally,}
\ceil{\frac{m}{T}}
&= \floor{\frac{n}{T}}+1
\hspace{1.32em}
\text{when }
m\in \left(\floor{\frac{n}{T}}T,n\right].
\end{align*}
Since $\PP{\BB{m}{p}\geq k}$ is nondecreasing in $m$ for constant $p$ and $k$, it follows that $\Ps(p,T,m)$ is maximized within each of these intervals of $m$ when we pick $m$ to be the largest integer in the corresponding interval.
Thus, given $n$, $p$, and $T$, we can find an optimal $m^*$ that maximizes $\Ps(p,T,m)$ over all $m$ from among $\ceil{\frac{n}{T}}$ candidates:
\begin{align}
\left\{\floor{T},\floor{2T},\ldots,\floor{\floor{\frac{n}{T}}T},n\right\}. \label{eq:CandidateMs}
\end{align}
For $m=\floor{kT}$, where $k\in\ZZ^+$, the corresponding probability of successful recovery is given by
\[
\Psm{\floor{kT}}
= \PP{\BB{\floor{kT}}{p}\geq k}.
\]
The difference between the probabilities of successful recovery for consecutive values of $k\in\ZZ^+$ can be written as
\begin{align*}
& \Delta(p,T,k)
\triangleq \Psm{\floor{(k+1)T}} - \Psm{\floor{kT}}
\\
&\;\; = \PP{\BB{\floor{(k+1)T}}{p}\geq k+1} - \PP{\BB{\floor{kT}}{p}\geq k}
\\
&\;\; = \hspace{-2em} \sum_{i=1}^{\min(\alpha_{k,T}-1,k)} \hspace{-1.5em} \PP{\BB{\floor{kT}}{p}=k-i} \cdot \PP{\BB{\alpha_{k,T}}{p}\geq i+1}
\\*[0.5em]
&\;\;\hspace{1.2em} - \PP{\BB{\floor{kT}}{p}=k} \cdot \PP{\BB{\alpha_{k,T}}{p}=0},
\end{align*}
where $\alpha_{k,T} \triangleq \floor{(k+1)T}-\floor{kT}$.
The above expression is obtained by comparing the branches of the probability tree for $\floor{kT}$ vs.~$\floor{(k+1)T}$ independent $\text{Bernoulli}(p)$ trials:
the first term describes unsuccessful events
(``\m{$\BB{\floor{kT}}{p}<k$}'') becoming successful
(``\m{$\BB{\floor{(k+1)T}}{p}\geq k+1$}'') after the additional $\alpha_{k,T}$ trials, while the second term describes successful events
(``\m{$\BB{\floor{kT}}{p}\geq k$}'') becoming unsuccessful
(``\m{$\BB{\floor{(k+1)T}}{p}<k+1$}'') after the additional $\alpha_{k,T}$ trials.
After further simplification, we arrive at
\begin{align}
& \Delta(p,T,k)
= p^{k} (1-p)^{\floor{(k+1)T}-k} \cdot\notag
\\*
&
{\scriptsize
\begin{array}{l}\displaystyle
\;\,\left\{ \sum_{i=1}^{\min\big(\alpha_{k,T}-1,k\big)} \sum_{j=i+1}^{\alpha_{k,T}} \binom{\floor{kT}}{k-i} \binom{\alpha_{k,T}}{j} \left(\frac{p}{1-p}\right)^{-i+j} \!\! - \binom{\floor{kT}}{k} \right\}\!.
\end{array}}
\label{eq:DeltaK}
\end{align}

The following theorem essentially provides a sufficient condition on $p$ and $T$ for \m{$\Delta(p,T,k)\geq 0$} for any \m{$k\in\ZZ^+$}, thereby eliminating all but the two largest candidate values for $m^*$ in \eqref{eq:CandidateMs}, i.e.,
\m{$m=\floor{\floor{\frac{n}{T}}T}$} and \m{$m=n$}, which correspond to a maximal spreading of the budget over (almost) all nodes (they are identical when \m{$\frac{n}{T}\in\ZZ^+$},
i.e., $T=n,\frac{n}{2},\frac{n}{3},\ldots$):

\begin{thm:theorem}
\label{thm:theorem:IndeptProbAccessSymmMaxSpread}
If
\begin{align}
(1-p)^{\floor{T}} + 2 \floor{T} p (1-p)^{\floor{T}-1} - 1 \leq 0, \label{eq:MaxCondition}
\end{align}
then either \xxsm{\floor{\floor{\frac{n}{T}}T}} or \xxsm{n}, which correspond to a maximal spreading of the budget, is an optimal symmetric allocation.
\end{thm:theorem}

\noindent
The following corollary restates Theorem~\ref{thm:theorem:IndeptProbAccessSymmMaxSpread} in a slightly weaker but more convenient form:

\begin{thm:corollary}
\label{thm:corollary:IndeptProbAccessSymmMaxSpread}
If \m{$p\geq\frac{4}{3\floor{T}}$},
then either \xxsm{\floor{\floor{\frac{n}{T}}T}} or \xxsm{n} is an optimal symmetric allocation.
\end{thm:corollary}

The following lemma mirrors Theorem~\ref{thm:theorem:IndeptProbAccessSymmMaxSpread} by providing a sufficient condition on $p$ and $T$ for \m{$\Delta(p,T,k)\leq 0$} for any \m{$k\in\ZZ^+$}, thereby eliminating all but the smallest candidate value for $m^*$ in \eqref{eq:CandidateMs}, i.e., \m{$m=\floor{T}$}, which corresponds to a minimal spreading of the budget (uncoded replication):

\begin{thm:lemma}
\label{thm:lemma:IndeptProbAccessSymmMinSpread1}
If \m{$T>1$}, and either
\begin{align}
T&=\frac{1}{p}\in\ZZ^+ \label{eq:MinCondition1}
\end{align}
or
\begin{align}
T<\frac{1}{p} \;\;\;\text{and}\;\;\; p\left(1-p\right)^{\ceil{T}-1} \leq \frac{1}{T}\left(1-\frac{1}{T}\right)^{\ceil{T}-1}, \label{eq:MinCondition2}
\end{align}
then \xxsm{\floor{T}} is an optimal symmetric allocation.
\end{thm:lemma}

\noindent
The following lemma restates Lemma~\ref{thm:lemma:IndeptProbAccessSymmMinSpread1} in a slightly weaker but more convenient form:

\begin{thm:lemma}
\label{thm:lemma:IndeptProbAccessSymmMinSpread2}
If \m{$p\leq\frac{2}{\ceil{T}}-\frac{1}{T}$},
then \xxsm{\floor{T}} is an optimal symmetric allocation.
\end{thm:lemma}

\begin{figure}
\centering
\includegraphics[width=0.47\textwidth]{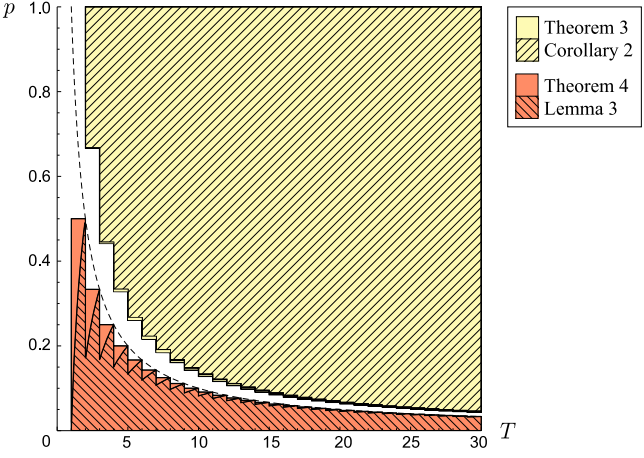}
\caption{Plot of access probability $p$ against budget $T$, showing regions of \m{$(T,p)$} over which the sufficient conditions of the theorems are satisfied.
The black dashed curve marks the points satisfying \m{$p=\frac{1}{T}$}.
Maximal spreading is optimal among symmetric allocations in the colored regions above the curve,
while minimal spreading (uncoded replication) is optimal among symmetric allocations in the colored regions below the curve.}
\label{fig:IndeptProbAccessOptSymmRegionPlot}
\end{figure}

\noindent
The following theorem expands the region covered by Lemma~\ref{thm:lemma:IndeptProbAccessSymmMinSpread2} by showing that \xxsm{\floor{T}} remains optimal between the ``peaks'' in Fig.~\ref{fig:IndeptProbAccessOptSymmRegionPlot}:

\begin{thm:theorem}
\label{thm:theorem:IndeptProbAccessSymmMinSpread}
If \m{$p\leq\frac{1}{\ceil{T}}$},
then \xxsm{\floor{T}}, which corresponds to a minimal spreading of the budget (uncoded replication), is an optimal symmetric allocation.
\end{thm:theorem}

Fig.~\ref{fig:IndeptProbAccessOptSymmRegionPlot} illustrates these results in the form of a region plot.
The theorems cover all choices of $p$ and $T$ except for the gap around \m{$p=\frac{1}{T}$}, which diminishes with increasing $T$.
Both minimal and maximal spreading of the budget may be suboptimal among symmetric allocations in this gap on either side of the curve \m{$p=\frac{1}{T}$}:
for example, when \m{$(n,p,T)=\left(10,\frac{9}{25},\frac{5}{2}\right)$}, for which \m{$p<\frac{1}{T}$}, the optimal symmetric allocation is \xxsm{\floor{2T}};
when \m{$(n,p,T)=\left(10,\frac{3}{5},\frac{12}{5}\right)$}, for which \m{$p>\frac{1}{T}$}, the optimal symmetric allocation is \xxsm{\floor{3T}}.
In general, for any budget \m{$T\geq 2$}, the optimal symmetric allocation changes from minimal spreading to maximal spreading eventually, as the access probability $p$ increases.
This transition, which is not necessarily sharp, appears to occur at around \m{$p=\frac{1}{T}$}.
Interestingly, when \m{$p=\frac{1}{T}$} exactly, we observe numerically that \xxsm{\floor{T}} is the optimal symmetric allocation for \emph{most} values of $T$;
the optimal symmetric allocation changes continually over the intervals
\[
1.5 \leq T < 2
\quad\text{and}\quad
2.5 \leq T \leq 2.8911,
\]
while \xxsm{\floor{2T}} is optimal for
\m{$3.5 \leq T \leq 3.5694$}.
These findings suggest that it may be difficult to specify an optimal symmetric allocation for values of $p$ and $T$ in the gap;
we can, however, restrict our search for an optimal symmetric allocation to the $\ceil{\frac{n}{T}}$ candidates given by \eqref{eq:CandidateMs}.

\section{\hspace{-0.5em} Access to a Random Fixed-Size Subset of Nodes}
\label{sec:RandFixedSizeSubset}

In the second variation of the storage allocation problem, we consider a data collector that accesses an \m{$r$-subset} of the $n$ nodes selected uniformly at random from the collection of all $\binom{n}{r}$ possible \m{$r$-subsets}, where $r$ is a given constant;
successful recovery occurs if and only if the total amount of data stored in the accessed nodes is at least $1$.
We seek an optimal allocation $\xxn$ of the budget $T$ that maximizes the probability of successful recovery, for a given choice of $n$, $r$, and $T$.
This optimization problem can be expressed as follows:
\begin{align}
& \PPP_2(n,r,T): \hspace{17em} \notag
\\*
& \hspace{1.8em} \underset{x_1,\ldots,x_n,\Ps}{\text{maximize}} \qquad \Ps \label{eq:P2ObjFunc}
\\*[0.5em]
& \hspace{1.9em} \text{subject to} \notag
\\*
& \hspace{3.6em} \sum_{\substack{\rr\subseteq\{1,\ldots,n\}:\\|\rr|=r}}
\frac{1}{\binom{n}{r}}
\cdot \II{\sum_{i\in\rr} x_i \geq 1}
\hspace{0em} \geq \hspace{0em}
\Ps \label{eq:P2Constraint1}
\\*
& \hspace{4.95em} \sum_{i=1}^{n} x_i \hspace{0em} \leq \hspace{0.5em} T \label{eq:P2Constraint2}
\\*
& \hspace{6.2em} x_i \hspace{0.5em} \geq \hspace{0.5em} 0 \hspace{2.05em} \forall\;\; i\in\nset. \label{eq:P2Constraint3}
\end{align}
The left-hand side of inequality~\eqref{eq:P2Constraint1} is just the recovery probability, expressed as the sum of the probabilities corresponding to the $r$-subsets $\rr$ that allow successful recovery.
The objective function~\eqref{eq:P2ObjFunc} is therefore equal to the recovery probability since $\Ps$ is maximized when \eqref{eq:P2Constraint1} holds with equality.
Inequality~\eqref{eq:P2Constraint2} expresses the budget constraint, and inequality~\eqref{eq:P2Constraint3} ensures that a nonnegative amount of data is stored in each node.
For the trivial budget \m{$T=1$}, the allocation \m{$(1,0,\ldots,0)$} is optimal;
for \m{$T\geq\frac{n}{r}$}, the allocation $\xxr$, which has the maximal recovery probability of $1$, is optimal.
Incidentally, computing the recovery probability of a \emph{given} allocation turns out to be \#P-complete:
\begin{thm:proposition}
\label{thm:proposition:RandFixedSizeSubsetPComplete}
Computing the recovery probability
\[
\sum_{\substack{\rr\subseteq\{1,\ldots,n\}:\\|\rr|=r}}
\frac{1}{\binom{n}{r}}
\cdot \II{\sum_{i\in\rr} x_i \geq 1}
\]
for a given allocation $\xxn$ and choice of $r$ is \#P-complete.
\end{thm:proposition}

An alternate way of formulating this problem is to minimize the budget $T$ required to achieve a desired recovery probability $\Ps$:
\begin{align*}
& \PPP'_2(n,r,\Ps): \hspace{17em}
\\*
& \hspace{1.8em} \underset{x_1,\ldots,x_n,T}{\text{minimize}} \qquad T
\\*[0.5em]
& \hspace{1.9em} \text{subject to the three constraints \eqref{eq:P2Constraint1}--\eqref{eq:P2Constraint3} of } \PPP_2(n,r,T).
\end{align*}
\begin{figure}
\centering
\subfloat[$(n,r)=(6,2)$]{%
\includegraphics[width=0.45\textwidth]{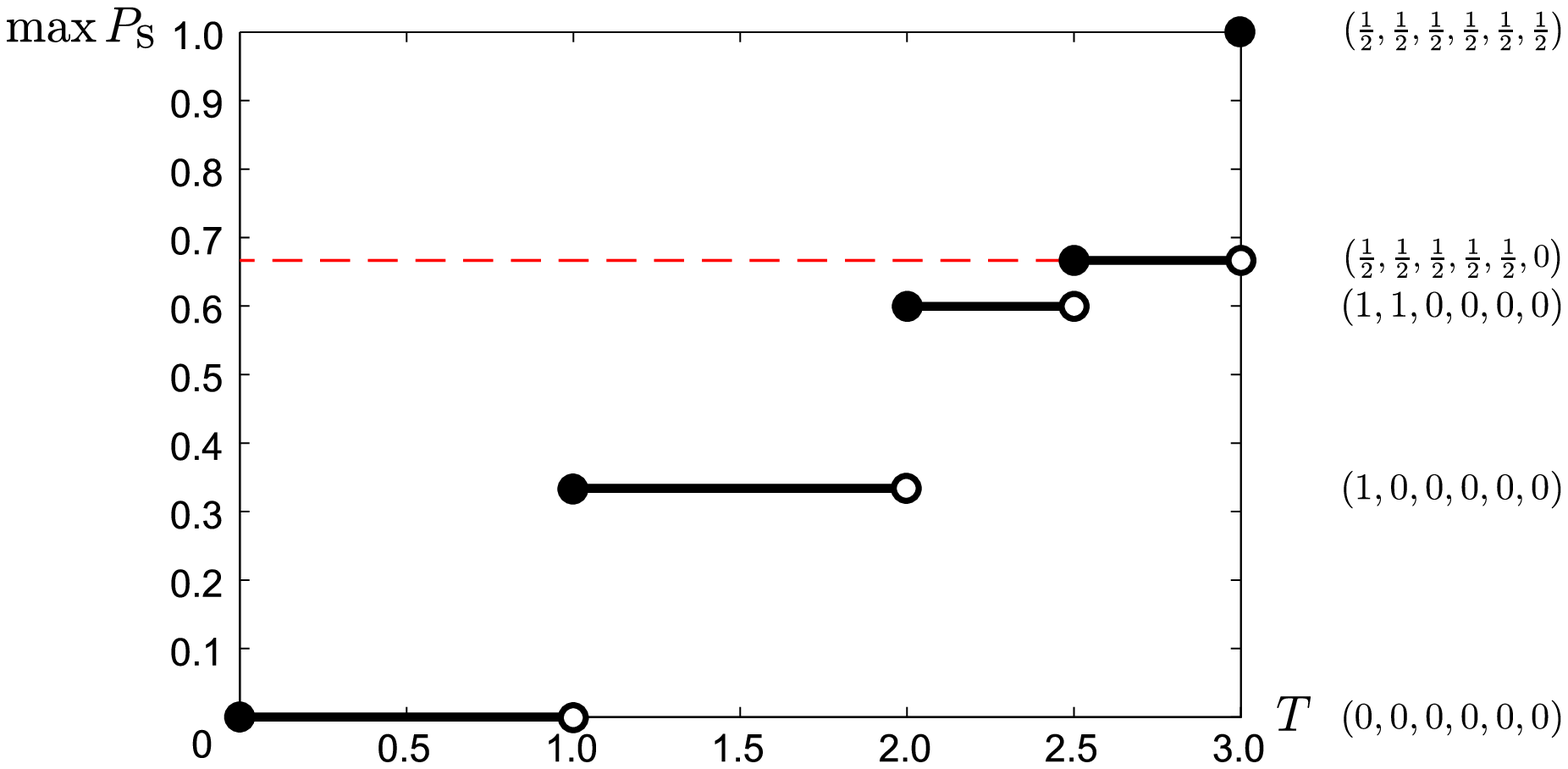}}
\\
\subfloat[$(n,r)=(5,3)$]{%
\includegraphics[width=0.45\textwidth]{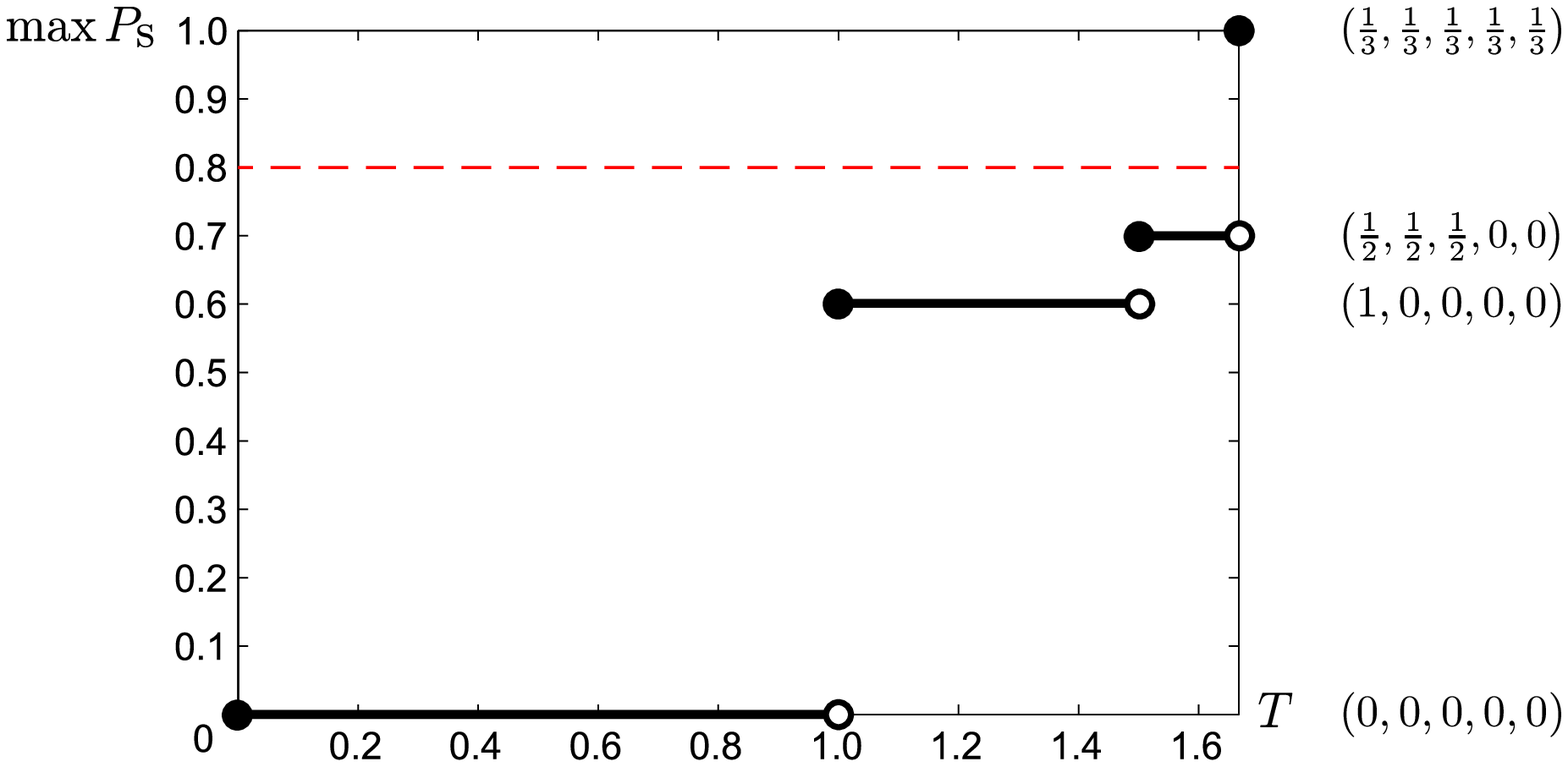}}
\caption{Plot of the optimal recovery probability \m{$\max\Ps$} against budget~$T$, for (a) \m{$(n,r)=(6,2)$} and (b) \m{$(n,r)=(5,3)$}.
The optimal allocation corresponding to each value of \m{$\max\Ps$} is given on the right-hand side of the plot.
In~(a), the red dashed line marks the threshold on $\Ps$ derived in Theorem~\ref{thm:theorem:RandFixedSizeSubsetIffDiv};
the allocation $\xxr$ is optimal for \m{$\PPP'_2(n,r,\Ps)$} if and only if the desired recovery probability $\Ps$ exceeds this threshold.
In~(b), the red dashed line marks the threshold on $\Ps$ derived in Theorem~\ref{thm:theorem:RandFixedSizeSubsetNotDiv};
the allocation $\xxr$ is optimal for \m{$\PPP'_2(n,r,\Ps)$} if $\Ps$ exceeds this threshold.}
\label{fig:RandFixedSizeSubsetPlots}
\end{figure}%
Fig.~\ref{fig:RandFixedSizeSubsetPlots} shows how the optimal recovery probability \m{$\max\Ps$} varies with the budget $T$, for two instances of \m{$(n,r)$}.
These plots are obtained by solving \m{$\PPP'_2(n,r,\Ps)$} for each possible value of $\Ps$.
We observe that when the budget $T$ drops below $\frac{n}{r}$, the optimal recovery probability \m{$\max\Ps$} is reduced by a significant margin below $1$.
In other words, if the desired recovery probability $\Ps$ in \m{$\PPP'_2(n,r,\Ps)$} is sufficiently high, then the optimal allocation is $\xxr$, which requires a budget of \m{$T=\frac{n}{r}$}.
In Section~\ref{sec:RandFixedSizeSubsetRegimeHighRecovProb}, we examine the optimality of this allocation for the high recovery probability regime.

\subsection{Regime of High Recovery Probability}
\label{sec:RandFixedSizeSubsetRegimeHighRecovProb}

Consider the optimization problem \m{$\PPP'_2(n,r,\Ps)$}.
We will demonstrate that the allocation $\xxr$ is optimal when the desired recovery probability $\Ps$ exceeds a specified threshold expressed in terms of $n$ and $r$.
Our results follow from the observation that successful recovery for certain combinations of \m{$r$-subsets} of nodes can impose a lower bound on the required budget $T$.
For example, given \m{$(n,r)=(4,2)$}, if successful recovery is to occur for $\{1,2\}$ and $\{3,4\}$, possibly among other \m{$r$-subsets} of nodes, then we have
\[
\sum_{i\in\{1,2\}} x_i \geq 1
\quad\text{and}\quad
\sum_{i\in\{3,4\}} x_i \geq 1,
\]
which would imply that the minimum budget $T$ must be at least $2$, since
\[
T\geq\sum_{i=1}^{4} x_i
= \sum_{i\in\{1,2\}} x_i + \sum_{i\in\{3,4\}} x_i
\geq 2.
\]
This observation is generalized by the following lemma:

\begin{thm:lemma}
\label{thm:lemma:RandFixedSizeSubsetCover}
Consider a set \m{$S\subseteq\nset$}, and $c$ subsets of $S$ given by \m{$\rr_j\subseteq S$}, \m{$j=1,\ldots,c$}.
If
\begin{align}
\sum_{i \in \rr_j} x_i \geq 1 \quad \forall\;\; j\in\{1,\ldots,c\}, \label{eq:inequalitiesm}
\end{align}
and each element in $S$ appears exactly \m{$b>0$} times among the $c$ subsets, i.e.,
\begin{align}
\sum_{j=1}^{c} \II{i\in\rr_j} = b \quad \forall\;\; i \in S, \label{eq:variablecount}
\end{align}
then
\[
\sum_{i \in S} x_i \geq \frac{c}{b}.
\]
\end{thm:lemma}

\begin{figure*}[!b]
\setcounter{tempeqcounter}{\value{equation}}
\setcounter{equation}{19} %
\hrulefill
\begin{align}
& P_{\text{S}}^{\textsf{UB}}
\triangleq\!
\max_{\ell\in\{0,1,\ldots,\floor{T}\}}
\overbrace{
1-(1-p)^{\ell}
}^{\text{(i)}}
\;+\;
\overbrace{
\II{\ell<\floor{T}}\cdot
(1-p)^{\ell}\cdot
\hspace{27.5em}}^{\text{(ii)}}
\notag
\\*[-0.5em]
& \hspace{15.2em}
\sum_{r=2}^{n-\ell}
\min\Bigg(
\underbrace{
\frac{r(T-\ell)}{n-\ell}
}_{\text{cf.~Lemma~\ref{thm:lemma:IndeptProbAccessUpperBound}}},\;
\underbrace{
1-\II{T-\ell<\frac{n-\ell}{r}}\cdot
\frac{\gcd(r,r')}{\alpha\gcd(r,r')+r'}
}_{\text{cf.~Theorems~\ref{thm:theorem:RandFixedSizeSubsetIffDiv} and \ref{thm:theorem:RandFixedSizeSubsetNotDiv}}}
\Bigg)\cdot
\PP{\BB{n-\ell}{p}=r}.
\label{eq:ImprovedUpperBound}
\end{align}
\setcounter{equation}{\value{tempeqcounter}}
\end{figure*}

We begin with the special case of probability-$1$ recovery, i.e., \m{$\Ps=1$}.
The resulting optimization problem is just a linear program with all $\binom{n}{r}$ possible \m{$r$-subset} constraints.

\begin{thm:lemma}
\label{thm:lemma:RandFixedSizeSubsetProb1}
If \m{$\Ps=1$}, then $\xxr$ is an optimal allocation.
\end{thm:lemma}

\noindent
When the desired recovery probability $\Ps$ is less than $1$, we can afford to drop \emph{some} of the \m{$r$-subset} constraints from this linear program
(recall that the recovery probability of an allocation is just the fraction of these $\binom{n}{r}$ constraints that are satisfied).
Our task is to determine how many such constraints can be dropped before the lower bound for $T$ obtained with the help of Lemma~\ref{thm:lemma:RandFixedSizeSubsetCover} falls below $\frac{n}{r}$, in which case the allocation $\xxr$ may no longer be optimal.
We do this by constructing collections of \m{$r$-subset} constraints that yield the required lower bound of $\frac{n}{r}$ for $T$, and counting how many \m{$r$-subset} constraints need to be removed from the linear program before no such collection remains.
Our answer depends on the divisibility of $n$ by $r$.

When $n$ is a multiple of $r$, we are able to state a necessary and sufficient condition on $\Ps$ for the allocation to be optimal:

\begin{thm:theorem}
\label{thm:theorem:RandFixedSizeSubsetIffDiv}
If $n$ is a multiple of $r$, then $\xxr$ is an optimal allocation if and only if
\[
\Ps > 1-\frac{r}{n}.
\]
\end{thm:theorem}

When $n$ is \emph{not} a multiple of $r$, we are only able to state a sufficient condition on $\Ps$ for the allocation to be optimal:

\begin{thm:theorem}
\label{thm:theorem:RandFixedSizeSubsetNotDiv}
If $n$ is not a multiple of $r$, then $\xxr$ is an optimal allocation if
\[
\Ps > 1-\frac{\gcd(r,r')}{\alpha\gcd(r,r')+r'},
\]
where $\alpha$ and $r'$ are uniquely defined integers satisfying
\[
n=\alpha\,r+r',\quad
\alpha\in\ZZ^+_0,\quad
r'\in\{r+1,\ldots,2r-1\}.
\]
\end{thm:theorem}

\noindent
However, if $n$ is a multiple of \m{$(n-r)$}, then this sufficient condition becomes necessary too:

\begin{thm:corollary}
\label{thm:corollary:RandFixedSizeSubsetIffDiffDiv}
If $n$ is a multiple of \m{$(n-r)$}, then $\xxr$ is an optimal allocation if and only if
\[
\Ps > \frac{r}{n}.
\]
\end{thm:corollary}

\noindent
Note that Corollary~\ref{thm:corollary:RandFixedSizeSubsetIffDiffDiv} allows us to solve \m{$\PPP_2(n,r,T)$} \emph{completely} when $n$ is a multiple of \m{$(n-r)$}:
for any \m{$T\in\left[1,\frac{n}{r}\right)$}, the allocation \m{$(1,0,\ldots,0)$} is optimal since it has a recovery probability of
$\frac{\binom{n-1}{r-1}}{\binom{n}{r}}=\frac{r}{n}$, i.e., exactly the threshold in Corollary~\ref{thm:corollary:RandFixedSizeSubsetIffDiffDiv};
higher recovery probabilities are not achievable unless \m{$T\geq\frac{n}{r}$}.

\begin{figure}
\centering
\includegraphics[width=0.47\textwidth]{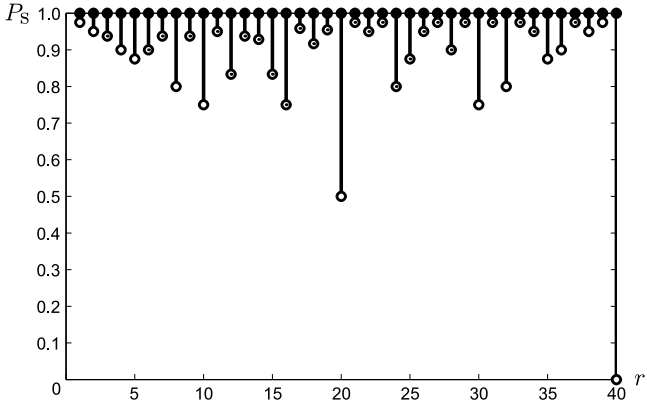}
\caption{Plot of the desired recovery probability $\Ps$ against the number of nodes accessed $r$, showing intervals of $\Ps$ over which the allocation $\xxr$ is optimal for \m{$\PPP'_2(n,r,\Ps)$}, for \m{$n=40$}.
A dotted circle marker denotes an endpoint that may not be tight, i.e., we have not demonstrated that the allocation is suboptimal everywhere outside the interval.}
\label{fig:RandFixedSizeSubsetOptPlot}
\end{figure}%
Fig.~\ref{fig:RandFixedSizeSubsetOptPlot} illustrates these results for an instance of $n$.

\addtocounter{equation}{1}
By combining the proof techniques of Lemma~\ref{thm:lemma:IndeptProbAccessUpperBound} and Theorems~\ref{thm:theorem:IndeptProbAccessMinSpreadOpt}, \ref{thm:theorem:RandFixedSizeSubsetIffDiv}, and \ref{thm:theorem:RandFixedSizeSubsetNotDiv}, we can derive the improved upper bound~$P_{\text{S}}^{\textsf{UB}}$, given by \eqref{eq:ImprovedUpperBound}
at the bottom of the page,
for the recovery probability of an optimal allocation in the independent probabilistic access model of Section~\ref{sec:IndeptProbAccess} (cf.~Lemma~\ref{thm:lemma:IndeptProbAccessUpperBound}).
Variables $\alpha$ and $r'$ are uniquely defined integers satisfying
\[
n-\ell=\alpha\,r+r',\quad
\alpha\in\ZZ^+_0,\quad
r'\in\{r,\ldots,2r-1\}.
\]
Parameter $\ell$ denotes the number of individual nodes that store at least a unit amount of data.
At least $\ell$ amount of data is stored in these \emph{complete} nodes, leaving the remaining budget of at most \m{$T-\ell$} to be allocated over the remaining \m{$n-\ell$} \emph{incomplete} nodes.
Term~(i) gives the probability of successful recovery from accessing at least one complete node, while term~(ii) gives an upper bound on the probability of successful recovery from accessing exactly $r\in$ \m{$\{2,\ldots,n-\ell\}$} incomplete nodes.

\section{Probabilistic Symmetric Allocations}
\label{sec:ProbAlloc}

In the third variation of the storage allocation problem, we consider the case where each of the $n$ nodes is selected by the source independently with probability
\m{$\min\!\left(\frac{\ell T}{n},1\right)$} to store $\frac{1}{\ell}$ amount of data, so that the expected total amount of storage used in the resulting \emph{symmetric} allocation is at most
\m{$n\cdot\frac{\ell T}{n}\cdot\frac{1}{\ell}$} $=T$,
the given budget.
The data collector subsequently accesses an \m{$r$-subset} of the $n$ nodes selected uniformly at random from the collection of all $\binom{n}{r}$ possible \m{$r$-subsets}, where $r$ is a given constant;
successful recovery occurs if and only if the total amount of data stored in the accessed nodes is at least $1$.
We seek an optimal probabilistic symmetric allocation of the budget $T$, specified by the value of parameter $\ell$, that maximizes the probability of successful recovery, for a given choice of $n$, $r$, and $T$.
Since successful recovery for a particular choice of $\ell$ occurs if and only if at least
\m{$\ceil{1\big/\left(\frac{1}{\ell}\right)}$} $=\ceil{\ell}$ out of the $r$ accessed nodes are nonempty, the corresponding probability of successful recovery can be written as
\[
\textstyle
\Ps(n,r,T,\ell) \triangleq
\PP{\BB{r\vphantom{\Big|}}{\min\!\left(\frac{\ell T}{n},1\right)}\geq\ceil{\ell\vphantom{\big|}}}.
\]
This optimization problem can therefore be expressed as follows:
\begin{align*}
& \PPP_3(n,r,T): \hspace{17em}
\\*
& \hspace{1.8em} \underset{\ell}{\text{maximize}} \hspace{1.5em} \textstyle
\PP{\BB{r\vphantom{\Big|}}{\min\!\left(\frac{\ell T}{n},1\right)}\geq\ceil{\ell\vphantom{\big|}}}
\\*[0.5em]
& \hspace{1.9em} \text{subject to} \hspace{1.4em}
\ell >0.
\end{align*}
For budget \m{$T\geq\frac{n}{r}$}, the choice of \m{$\ell=r$}, which yields a recovery probability of
\m{$\PP{\BB{r}{1}\geq r}$} $=1$, is optimal.

Observe that the recovery probability \m{$\Ps(n,r,T,\ell)$} is zero when \m{$\ell>r$}.
Furthermore, for fixed $n$, $r$, and $T$, the recovery probability is nondecreasing in $\ell$ within each of the unit intervals
\m{$(0,1]$}, \m{$(1,2]$}, \m{$(2,3]$}, $\ldots$, since as $\ell$ increases within each interval, $\ceil{\ell}$ remains constant while
\m{$\min\!\left(\frac{\ell T}{n},1\right)$}
either increases or remains constant at $1$.
Thus, given $n$, $r$, and $T$, we can find an optimal $\ell^*$ from among $r$ candidates:
\begin{align}
\big\{1,2,\ldots,r\big\}. \label{eq:CandidateLs}
\end{align}

\begin{figure}
\centering
\includegraphics[width=0.49\textwidth]{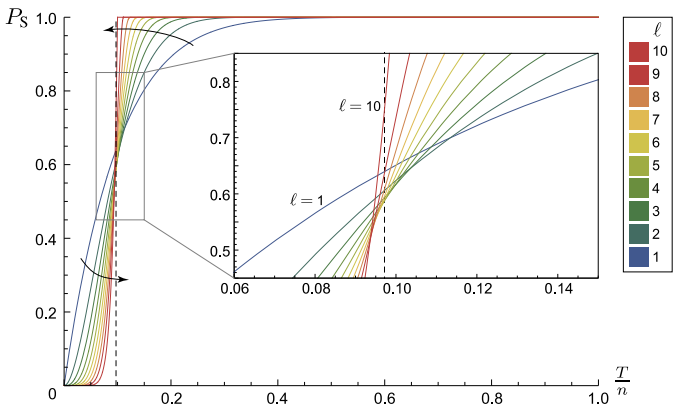}
\caption{Plot of recovery probability $\Ps$ against budget-per-node $\frac{T}{n}$ for each choice of parameter
\m{$\ell\in\{1,2,\ldots,r\}$}, for \m{$r=10$}.
Parameter $\ell$ controls how much the budget is spread in the probabilistic symmetric allocation;
specifically, each of the $n$ nodes is selected by the source independently with probability
\m{$\min\!\big(\frac{\ell T}{n},1\big)$}
to store $\frac{1}{\ell}$ amount of data.
Arrows indicate the direction of increasing $\ell$.
The black dashed line marks the threshold on $\frac{T}{n}$ derived in Theorem~\ref{thm:theorem:ProbAllocMaxSpreadOptT};
maximal spreading (\m{$\ell=r$}) is optimal for any $\frac{T}{n}$ greater than or equal to this threshold.}
\label{fig:ProbAllocPlotR10}
\end{figure}%
Fig.~\ref{fig:ProbAllocPlotR10}, which compares the performance of different probabilistic symmetric allocations over different budgets for an instance of $r$, suggests that there are two distinct phases pertaining to the optimal choice of $\ell$:
when the budget is below a certain threshold, the choice of \m{$\ell=1$}, which corresponds to a minimal spreading of the budget (uncoded replication), is optimal;
when the budget exceeds that same threshold, the choice of \m{$\ell=r$}, which corresponds to a maximal spreading of the budget, becomes optimal.
This observation echoes our findings on the allocation and access models of the preceding sections, namely that minimal spreading \m{($\ell=1$)} is optimal for sufficiently small budgets, while maximal spreading \m{($\ell=r$)} is optimal for sufficiently large budgets.
However, we note two important distinctions in contrast to the previous models.
First, the recovery probability for a probabilistic symmetric allocation in this model is a \emph{continuous} nondecreasing function of the given budget;
there are no ``jumps'' from one discrete value to the next.
Second, our empirical computations suggest that the phase transition from the optimality of minimal spreading to that of maximal spreading in this model is \emph{sharp};
the other intermediate values of \m{$\ell\in\{2,\ldots,r-1\}$} never perform better than both \m{$\ell=1$} and \m{$\ell=r$} simultaneously.

In Section~\ref{sec:ProbAllocOptMaxSpreading}, we shall demonstrate that the choice of \m{$\ell=r$}, which corresponds to a maximal spreading of the budget, is indeed optimal when the given budget $T$ is sufficiently large, or equivalently, when a sufficiently high recovery probability is achievable.

\subsection{Optimality of Maximal Spreading}
\label{sec:ProbAllocOptMaxSpreading}

Assume that \m{$r\geq 2$}.
As noted earlier, the choice of \m{$\ell=r$}, which corresponds to a maximal spreading of the budget, is optimal for any \m{$T\geq\frac{n}{r}$} because it yields the maximal recovery probability of $1$.
The following lemma provides an upper bound for the recovery probabilities corresponding to the \emph{other} candidate values for $\ell^*$ in \eqref{eq:CandidateLs} at the critical budget \m{$T=\frac{n}{r}$}:

\begin{thm:lemma}
\label{thm:lemma:ProbAllocUpperBoundOtherL}
\hspace{-0.29em}
The probability of successful recovery \m{$\Ps(n,r,T,\ell)$} at \m{$T=\frac{n}{r}$} is at most $\frac{3}{4}$ for any
\m{$\ell\in\{1,2,\ldots,r-1\}$}.
\end{thm:lemma}

\noindent
Such an upper bound allows us to derive a sufficient condition for the optimality of \m{$\ell=r$}, by making use of the fact that the recovery probability \m{$\Ps(n,r,T,\ell)$} is a nondecreasing function of the budget $T$.
The following theorem shows that the choice of \m{$\ell=r$} is optimal when the budget $T$ is at least a specified threshold expressed in terms of $n$ and $r$:

\begin{thm:theorem}
\label{thm:theorem:ProbAllocMaxSpreadOptT}
If
\[
T \geq \frac{n}{r} \left(\frac{3}{4}\right)^{\frac{1}{r}},
\]
then the choice of \m{$\ell=r$}, which corresponds to a maximal spreading of the budget, is optimal.
\end{thm:theorem}

\noindent
The following corollary states an equivalent result in terms of the achievable recovery probability;
it demonstrates the optimality of \m{$\ell=r$} in the high recovery probability regime:

\begin{thm:corollary}
\label{thm:corollary:ProbAllocMaxSpreadOptPs}
If a probability of successful recovery of at least $\frac{3}{4}$ is achievable for the given $n$, $r$, and $T$, then the choice of \m{$\ell=r$} is optimal.
\end{thm:corollary}

\begin{figure}
\centering
\includegraphics[width=0.48\textwidth]{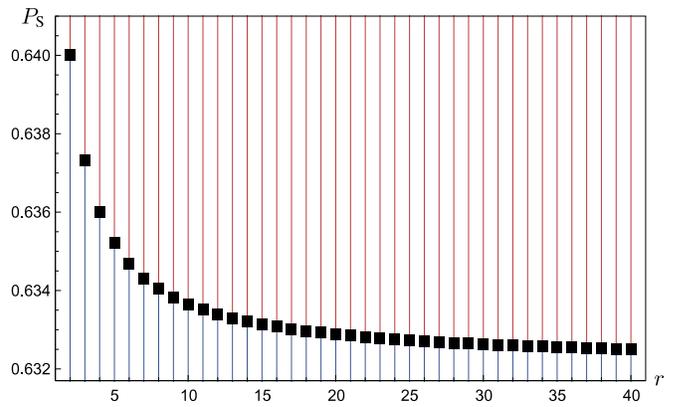}
\caption{Plot of recovery probability $\Ps$ against the number of nodes accessed $r$, indicating the value of $\Ps$ at which the optimal choice of parameter $\ell$ changes from $1$ to $r$, for each given value of $r$.
Specifically, if it is possible to achieve a recovery probability $\Ps$ above the square marker, then maximal spreading (\m{$\ell=r$}) is optimal;
otherwise, minimal spreading or uncoded replication (\m{$\ell=1$}) is optimal.
Observe that the critical value of $\Ps$ for \m{$r=10$} (which is approximately $0.633652$) corresponds to the intersection point of the curves for \m{$\ell=1$} and \m{$\ell=10$} in Fig.~\ref{fig:ProbAllocPlotR10}.}
\label{fig:ProbAllocOptPlot}
\end{figure}%
Fig.~\ref{fig:ProbAllocOptPlot} describes the optimal choice of $\ell$ for different values of $r$.
We observe that the gap between the threshold of $\frac{3}{4}$ derived in Corollary~\ref{thm:corollary:ProbAllocMaxSpreadOptPs} and the actual critical value of $\Ps$ indicated in the plot appears to be no more than $0.12$.

\section{Conclusion and Future Work}

We examined the problem of allocating a given total storage budget in a distributed storage system for maximum reliability.
Three variations of the problem were studied in detail, and we are able to specify the optimal allocation or optimal symmetric allocation for a variety of cases.
Although the exact optimal allocation is difficult to find in general, our results suggest a simple heuristic for achieving reliable storage:
\emph{when the budget is small, spread it minimally;
when the budget is large, spread it maximally.}
In other words, coding is unnecessary when the budget is small, but is beneficial when the budget is large.

The work in this paper can be extended in several directions.
We can impose additional system design constraints on the model;
one practical example is the application of a tighter per-node storage constraint $x_i\leq c_i<1$.
The independent probabilistic access model of Section~\ref{sec:IndeptProbAccess} can be naturally generalized to the case of nonuniform access probabilities $p_i$ for individual nodes.
It would also be interesting to find reliable allocations for specific codes with desirable encoding or decoding properties, e.g., sparse codes that offer efficient algorithms (see, e.g., \cite{dl:dimakis05ubiquitous,dl:kamra06growth,dl:lin07data,dl:aly08fountain}).
A related problem would be to construct such codes that work well under different allocations.
Another set of interesting problems involves the application of richer access models;
for instance, we can introduce a network topology to a set of storage nodes and data collectors, and allow each data collector to access only the nodes close to it.
More generally, we can assign different priorities to each node for data storage and access, so as to reflect the costs of storing data in the node and communicating with it.

\appendix[Proofs of Theorems]

\begin{IEEEproof}[Proof of Proposition~\ref{thm:proposition:IndeptProbAccessPHard}]
We note that the computational complexity of this problem was well understood in the Berkeley meetings \cite{dl:karppersonal} and is by no means a major contribution in this paper.
We present the detailed proofs here for completeness.

Consider an allocation $\xxn$ where each $x_i$ is a nonnegative rational number.
The problem of computing the recovery probability of this allocation for the special case of \m{$p=\frac{1}{2}$}, for which
\m{$p^{|\rr|}(1-p)^{n-|\rr|}=\left(\frac{1}{2}\right)^{n}$} for \emph{any} subset \m{$\rr\subseteq\nset$},
is equivalent to the counting version of the following decision problem (which happens to be polynomial-time solvable):

\begin{thm:definition}
\textsc{Largest Subset Sum} (\textsc{LSS})
\\
\textit{Instance}:
Finite $n$-vector \m{$\left(a_1,\ldots,a_n\right)$} with \m{$a_i\in\ZZ^+_0$}, and file size \m{$d\in\ZZ^+$}, where all $a_i$ and $d$ can be written as decimal numbers of length at most $\ell$.
\\
\textit{Question}:
Is there a subset \m{$\rr\subseteq\nset$} that satisfies
\m{$\sum_{i\in\rr} a_i\geq d$}?
\end{thm:definition}

Note that the allocation and file size have been scaled so that the problem parameters are all integers.
We will proceed to show that the counting problem \textsc{\#LSS} is \#P-complete;
this would in turn establish the \#P-hardness of computing the recovery probability for an arbitrary value of $p$.

The index set $\rr$ can be represented as an $n$-vector of bits.
Using this representation of $\rr$ as the certificate, it is easy to see that the binary relation corresponding to \textsc{\#LSS} is both polynomially balanced (since the size of each certificate is $n$), and polynomial-time decidable (since the inequality can be verified in \m{$O(n\ell)$} time for each certificate).
It therefore follows that \textsc{\#LSS} is in \#P.

To show that \textsc{\#LSS} is also \#P-hard, we describe a polynomial-time Turing reduction of the \#P-complete problem \textsc{\#3SAT} \cite{dl:valiant79complexity} to \textsc{\#LSS}.
Our approach is similar to the standard method of reducing \textsc{3SAT} to \textsc{Subset Sum} (see, e.g., \cite{dl:sipser06computation}).
Let $\phi$ be the Boolean formula in the given \textsc{\#3SAT} instance;
denote its $m$ variables by \m{$v_1,\ldots,v_m$},
and $k$ clauses by \m{$C_1,\ldots,C_k$}.
To count the number of satisfying truth assignments for $\phi$, we construct a \textsc{\#LSS} instance with the help of Table~\ref{tbl:IndeptProbAccessPHardProof}, whose entries are $0$, $1$, $2$, or $3$ (all blank entries are \m{$0$'s}).
\begin{table}
\caption{%
Constructing a \textsc{\#LSS} instance for a given \textsc{\#3SAT} instance}
\label{tbl:IndeptProbAccessPHardProof}
\centering
{\fontsize{7}{8.4}\selectfont
\noindent\begin{tabular}{@{\hspace{0.2em}}c@{\hspace{0.2em}}|%
c@{\hspace{0.4em}}c@{\hspace{0.4em}}c@{\hspace{0.4em}}c@{\hspace{0.4em}}|%
c@{\hspace{0.4em}}c\dashvertical%
c@{\hspace{0.4em}}c\dashvertical%
c\dashvertical%
c@{\hspace{0.4em}}c}
& $v_1$ & $v_2$ & $\cdots$ & $v_m$ &
\multicolumn{2}{c\dashvertical}{$C_1$} &
\multicolumn{2}{c\dashvertical}{$C_2$} &
$\cdots$ &
\multicolumn{2}{c}{$C_k$}
\\ \hline
$v_1$ & $1$ & & & & $\II{v_1{\in}C_1}$ & $0$ & $\II{v_1{\in}C_2}$ & $0$ & $\cdots$ & $\II{v_1{\in}C_k}$ & $0$
\\
$\overline{v_1}$ & $1$ & & & & $\II{\overline{v_1}{\in}C_1}$ & $0$ & $\II{\overline{v_1}{\in}C_2}$ & $0$ & $\cdots$ & $\II{\overline{v_1}{\in}C_k}$ & $0$
\\ \dashhorizontal
$v_2$ & & $1$ & & & $\II{v_2{\in}C_1}$ & $0$ & $\II{v_2{\in}C_2}$ & $0$ & $\cdots$ & $\II{v_2{\in}C_k}$ & $0$
\\
$\overline{v_2}$ & & $1$ & & & $\II{\overline{v_2}{\in}C_1}$ & $0$ & $\II{\overline{v_2}{\in}C_2}$ & $0$ & $\cdots$ & $\II{\overline{v_2}{\in}C_k}$ & $0$
\\ \dashhorizontal
$\vdots$ & & & $\ddots$ & & $\vdots$ & $\vdots$ & $\vdots$ & $\vdots$ & $\vdots$ & $\vdots$ & $\vdots$
\\ \dashhorizontal
$v_m$ & & & & $1$ & $\II{v_m{\in}C_1}$ & $0$ & $\II{v_m{\in}C_2}$ & $0$ & $\cdots$ & $\II{v_m{\in}C_k}$ & $0$
\\
$\overline{v_m}$ & & & & $1$ & $\II{\overline{v_m}{\in}C_1}$ & $0$ & $\II{\overline{v_m}{\in}C_2}$ & $0$ & $\cdots$ & $\II{\overline{v_m}{\in}C_k}$ & $0$
\\ \hline
\multirow{3}{*}{$C_1$} & & & & & $0$ & $1$ & & & & &
\\
& & & & & $1$ & $1$ & & & & &
\\
& & & & & $2$ & $1$ & & & & &
\\ \dashhorizontal
\multirow{3}{*}{$C_2$} & & & & & & & $0$ & $1$ & & &
\\
& & & & & & & $1$ & $1$ & & &
\\
& & & & & & & $2$ & $1$ & & &
\\ \dashhorizontal
$\vdots$ & & & & & & & & & $\ddots$ & &
\\ \dashhorizontal
\multirow{3}{*}{$C_k$} & & & & & & & & & & $0$ & $1$
\\
& & & & & & & & & & $1$ & $1$
\\
& & & & & & & & & & $2$ & $1$
\\ \noalign{\hrule height 1pt}
$d$ & $1$ & $1$ & $\cdots$ & $1$ & $3$ & $1$ & $3$ & $1$ & $\cdots$ & $3$ & $1$
\end{tabular}
} %
\end{table}
The entries of the $n$-vector for the \textsc{\#LSS} instance are given by the first \m{$(2m+3k)$} rows of the table;
the file size $d$ is given by the last row of the table.
Each entry $a_i$, \m{$i\in\{1,\ldots,2m+3k\}$}, as well as $d$, is a positive integer with at most \m{$(m+2k)$} decimal digits.
Observe that the set of satisfying truth assignments for $\phi$ can be put in a one-to-one correspondence with the collection of subsets \m{$\rr\subseteq\{1,\ldots,2m+3k\}$} that satisfy \m{$\sum_{i\in\rr} a_i=d$};
for each \m{$i\in\{1,\ldots,m\}$}, we have \m{``$v_i$'' $\in\rr$} if and only if \m{$v_i=$ \textsc{true}}, and \m{``$\overline{v_i}$'' $\in\rr$} if and only if \m{$v_i=$ \textsc{false}}.
Therefore, if \m{$f\big((a_1,\ldots,a_n),d\big)$} is a subroutine for computing \textsc{\#LSS}, then the number of satisfying truth assignments can be computed by calling $f$ twice:
first with $d$ taking the value as prescribed above, and second with $d$ taking the prescribed value \emph{plus one}.
The difference between the outputs from the two subroutine calls is equal to the number of distinct subsets $\rr$ that satisfy \m{$\sum_{i\in\rr} a_i=d$}, which is equal to the number of satisfying truth assignments for $\phi$.
Finally, we note that this is indeed a polynomial-time Turing reduction since the table can be populated in \m{$O\left(m^2 k^2\right)$} simple steps, and the subroutine $f$ is called exactly twice.
\end{IEEEproof}

\begin{IEEEproof}[Proof of Lemma~\ref{thm:lemma:IndeptProbAccessUpperBound}]
Consider a feasible allocation $\xxn$;
we have \m{$\sum_{i=1}^{n} x_i \leq T$}, where
\m{$x_i \geq 0$}, \m{$i=1,\ldots,n$}.
Let $S_r$ denote the number of \m{$r$-subsets} of \m{$\{x_1,\ldots,x_n\}$} that have a sum of at least $1$, where \m{$r\in\nset$}.
By conditioning on the number of nodes accessed by the data collector, the probability of successful recovery for this allocation can be written as
\begin{align}
& \PP{\text{successful recovery}} \notag
\\
&\hspace{-0.2em} = \hspace{-0.2em}
\sum_{r=1}^{n} \hspace{-0.5em}
\begin{array}{l}
\PP{\text{successful recovery}\,|\,\text{exactly $r$ nodes were accessed}} \cdot
\\
\hspace{8.55em} \PP{\text{exactly $r$ nodes were accessed}}
\end{array} \notag
\\
&\hspace{-0.2em} = \hspace{-0.2em}
\sum_{r=1}^{n} \frac{S_r}{\binom{n}{r}} \cdot \PP{\BB{n}{p}=r}. \label{eq:OptimalBound}
\end{align}
We proceed to find an upper bound for $S_r$.
For a given $r$, we can write $S_r$ inequalities of the form \[
x'_1+\cdots+x'_r \geq 1.
\]
Summing up these $S_r$ inequalities produces an inequality of the form
\[
a_1 x_1 + \cdots + a_n x_n \geq S_r.
\]
Since each $x_i$ belongs to exactly $\binom{n-1}{r-1}$ distinct \m{$r$-subsets} of \m{$\left\{x_1,\ldots,x_n\right\}$}, it follows that \m{$0 \leq a_i \leq\binom{n-1}{r-1}$}, $i=1,\ldots,n$.
Therefore,
\begin{align*}
& S_r
\leq a_1 x_1 + \cdots + a_n x_n
\\*
&\hspace{7em}
\leq \binom{n-1}{r-1} \sum_{i=1}^{n} x_i
\leq \binom{n-1}{r-1} T.
\end{align*}
Since $S_r$ is also at most $\binom{n}{r}$, i.e., the total number of \m{$r$-subsets}, we have
\[
S_r \leq \min\left(\binom{n-1}{r-1}T, \binom{n}{r}\right).
\]
Substituting this bound into \eqref{eq:OptimalBound} completes the proof.
\end{IEEEproof}

\begin{IEEEproof}[Proof of Theorem~\ref{thm:theorem:IndeptProbAccessMaxSpreadSubopt}]
The suboptimality gap for the symmetric allocation \xxsm{n} is at most the difference between its recovery probability~\eqref{eq:MaxSpreadRecoveryProb} and the upper bound~\eqref{eq:RecoveryProbUpperBound} from Lemma~\ref{thm:lemma:IndeptProbAccessUpperBound} for the optimal recovery probability.
This difference is given by
\begin{align*}
&\quad \sum_{r=1}^{\ceil{\frac{n}{T}}-1} \frac{rT}{n} \binom{n}{r} p^{r} (1-p)^{n-r}
\\
&= T\;\sum_{r=1}^{\ceil{\frac{n}{T}}-1} \binom{n-1}{r-1} p^{r} (1-p)^{n-r}
\\
&= p\,T\;\sum_{r=1}^{\ceil{\frac{n}{T}}-1} \binom{n-1}{r-1} p^{r-1} (1-p)^{(n-1)-(r-1)}
\\
&= p\,T\;\sum_{\ell=0}^{\ceil{\frac{n}{T}}-2} \binom{n-1}{\ell} p^{\ell} (1-p)^{(n-1)-\ell}
\\
&= p\,T\;\PP{\displaystyle\BB{n-1}{p}\leq\ceil{\frac{n}{T}}-2}
\;\triangleq\;
\delta(n,p,T),
\end{align*}
as required.
Assuming now that \m{$p>\frac{1}{T}$}, we have
\begin{align}
\delta(n,p,T)
&\leq  p\,T\;\PP{\displaystyle\BB{n-1}{p}\leq\frac{n-1}{T}} \label{eq:BinomialIneq}
\\
&= p\,T\;\PP{\displaystyle\BB{n-1}{p}\leq\frac{1}{pT}(n-1)p} \notag
\\
&\leq p\,T\;\exp\left(-\frac{(n-1)p}{2}\left(1-\frac{1}{pT}\right)^2\right). \label{eq:ChernoffIneq}
\end{align}
Inequality~\eqref{eq:BinomialIneq} follows from the fact that
\[
\ceil{\frac{n}{T}}-2\;
<\; \frac{n}{T}+1-2\;
<\; \frac{n}{T}-\frac{1}{T}.
\]
Inequality~\eqref{eq:ChernoffIneq} follows from the observation that \m{$\frac{1}{pT}\in(0,1)$}, and the subsequent application of the Chernoff bound for deviation below the mean of the binomial distribution (see, e.g., \cite{dl:mitzenmacher05probability}).
For fixed $p$ and $T$, this upper bound approaches zero as $n$ goes to infinity.
\end{IEEEproof}

\begin{IEEEproof}[Proof of Theorem~\ref{thm:theorem:IndeptProbAccessMinSpreadOpt}]
We compare the recovery probability of \xxsm{\floor{T}} to an upper bound for the recovery probabilities of all other allocations.

Suppose that \m{$1<T<n$}.
Recall from \eqref{eq:MinSpreadRecoveryProb} that the probability of successful recovery for \xxsm{\floor{T}} is given by
\[
P_1(p,T) \triangleq 1 - (1-p)^{\floor{T}}.
\]

Consider a feasible allocation $\xxn$;
we have \m{$\sum_{i=1}^{n} x_i \leq T$}, where \m{$x_i\geq 0$}, \m{$i=1,\ldots,n$}.
Let $\ell$ be the number of individual nodes in this allocation that store at least a unit amount of data;
for brevity, we refer to these nodes as being \emph{complete}.
It follows from the budget constraint that the number of complete nodes $\ell\in$ \m{$\{0,1,\ldots,\floor{T}\}$}.
When \m{$\ell=\floor{T}$}, the allocation has a recovery probability identical to \m{$P_1(p,T)$}.
Now, assuming that $\ell\in$ \m{$\{0,1,\ldots,\floor{T}-1\}$}, successful recovery can occur in two ways:
\begin{enumerate}
\item\label{thm:IndeptProbAccessMinSpreadOpt:Case1}
when the accessed subset contains one or more complete nodes, which occurs with probability \m{$1-(1-p)^{\ell}$}, or
\item\label{thm:IndeptProbAccessMinSpreadOpt:Case2}
when the accessed subset contains no complete nodes but has a sum of at least $1$.
\end{enumerate}
In case~\ref{thm:IndeptProbAccessMinSpreadOpt:Case2}, the accessed subset would consist of two or more \emph{incomplete} nodes.
Using the argument in the proof of Lemma~\ref{thm:lemma:IndeptProbAccessUpperBound}, we can show that there are at most
\[
\min\left(\binom{n-\ell-1}{r-1}(T-\ell),\binom{n-\ell}{r}\right)
\]
\m{$r$-subsets} of incomplete nodes whose sum is at least $1$, since the total amount of data stored over the \m{$n-\ell$} incomplete nodes is at most \m{$T-\ell$}.
It follows then that the recovery probability for a feasible allocation with exactly $\ell\in$ \m{$\{0,1,\ldots,\floor{T}-1\}$} complete nodes is at most
\begin{align*}
& P_2(n,p,T,\ell) \triangleq 1 - (1-p)^{\ell} + (1-p)^{\ell} \cdot
\\*
& \quad\qquad \sum_{r=2}^{n-\ell} \min\left(\frac{T-\ell}{n-\ell}\cdot r,1\right) \binom{n-\ell}{r} p^r (1-p)^{n-\ell-r}.
\end{align*}

Thus,
\[
P_1(p,T) \geq P_2(n,p,T,\ell)
\]
for all $\ell\in$ \m{$\{0,1,\ldots,\floor{T}-1\}$} is a sufficient condition for \xxsm{\floor{T}} to be an optimal allocation.
After further simplification of this inequality, we arrive at inequality \eqref{eq:IndeptProbAccessMinSpreadOptIneq} as required.
\end{IEEEproof}

\begin{IEEEproof}[Proof of Corollary~\ref{thm:corollary:IndeptProbAccessMinSpreadOpt}]
Suppose that \m{$1<T<n$}.
We will show that the sufficient condition of Theorem~\ref{thm:theorem:IndeptProbAccessMinSpreadOpt} is satisfied for any \m{$p\leq\frac{2}{\left(n-\floor{T}\right)^2}$}.
Note that when \m{$n-\floor{T}=1$}, or equivalently \m{$T\in[n-1,n)$}, we have to show that \xxsm{\floor{T}} is an optimal allocation for \emph{any} $p$, i.e., in the interval $(0,1)$.

First, observe that the summation term in inequality~\eqref{eq:IndeptProbAccessMinSpreadOptIneq} is always nonnegative, i.e.,
\[
\sum_{r=2}^{\ceil{\frac{n-\ell}{T-\ell}}-1} \left(1-\frac{T-\ell}{n-\ell}\cdot r\right) \binom{n-\ell}{r} \left(\frac{p}{1-p}\right)^r \geq 0,
\]
since for any $r\in$ \m{$\left\{2,\ldots,\ceil{\frac{n-\ell}{T-\ell}}-1\right\}$}
and $\ell\in$ \m{$\{0,1,\ldots,\floor{T}-1\}$}, we have
\[
r\leq\ceil{\frac{n-\ell}{T-\ell}}-1
\Longleftrightarrow
r<\frac{n-\ell}{T-\ell}
\Longleftrightarrow
1-\frac{T-\ell}{n-\ell}\cdot r>0.
\]
Therefore, a simpler but weaker sufficient condition for \xxsm{\floor{T}} to be an optimal allocation is
\begin{align*}
& 1 - (1-p)^{\floor{T}-n} + \left(n-\left(\floor{T}-1\right)\right) \left(\frac{p}{1-p}\right) \geq 0 \\
\Longleftrightarrow
& 1 + \left(n-\floor{T}\right) p - (1-p)^{1-\left(n-\floor{T}\right)} \geq 0,
\end{align*}
which is an inequality in only two variables $p$ and \m{$s\triangleq n-\floor{T}$}, where $s\in$ \m{$\{1,\ldots,n-1\}$}.
When \m{$s=1$}, or equivalently \m{$T\in[n-1,n)$}, this inequality is satisfied for any \m{$p\in(0,1)$}, as required.
Defining the function
\[
f(s,p) \triangleq 1 + s\,p - (1-p)^{1-s},
\]
it suffices to show that \m{$f(s,p)\geq 0$} for any \m{$s\in\ZZ^+$}, \m{$s\geq 2$}, and \m{$p\in\left(0,\frac{2}{s^2}\right]$}.
We do this by demonstrating that for any \m{$s\in\ZZ^+$}, \m{$s\geq 2$}, the function \m{$f(s,p)$} is concave in $p$ on the interval \m{$p\in\left(0,\frac{2}{s^2}\right]$}, and is nonnegative at both endpoints, i.e., \m{$f(s,p{=}0)\geq 0$} and \m{$f\left(s,p{=}\frac{2}{s^2}\right)\geq 0$}.

The second-order partial derivative of \m{$f(s,p)$} wrt $p$ is given by
\[
\frac{\partial^2}{\partial p^2} f(s,p) = -s(s-1)(1-p)^{-1-s}.
\]
Since \m{$\frac{\partial^2}{\partial p^2} f(s,p)<0$} for any \m{$s\in\ZZ^+$}, \m{$s\geq 2$}, and \m{$p\in\left(0,\frac{2}{s^2}\right]$}, it follows that the function \m{$f(s,p)$} is concave in $p$ on the interval \m{$p\in\left(0,\frac{2}{s^2}\right]$} for any \m{$s\in\ZZ^+$}, \m{$s\geq 2$}.

Suppose that \m{$s\in\ZZ^+$}, \m{$s\geq 2$}.
Clearly, \m{$f(s,p{=}0)=0$}.
To show that \m{$f\left(s,p{=}\frac{2}{s^2}\right)\geq 0$}, we define the function
\[
g(s) \triangleq \ln\left(1+\frac{2}{s}\right) + (s-1) \ln\left(1-\frac{2}{s^2}\right),
\]
and show that \m{$g(s)\geq 0$} for any \m{$s\in\ZZ^+$}, \m{$s\geq 2$}.
Direct evaluation of the function gives us \m{$g(s{=}2)=0$}, and \m{$g(s{=}3)=$} \m{$\ln\frac{5}{3}-2\ln\frac{9}{7}>0$}.
For \m{$s\geq 4$}, we consider the derivatives of $g(s)$:
\begin{align*}
g'(s) &= \frac{1}{s} + \frac{1}{s+2} - \frac{2(s-2)}{s^2-2} + \ln\left(1-\frac{2}{s^2}\right), \\
g''(s) &= \frac{8\left(s^3-s^2-6s-2\right)}{s^2 (s+2)^2 \left(s^2-2\right)^2}.
\end{align*}
Since \m{$g''(s)\geq 0$} for any \m{$s\geq 4$}, and \m{$\lim_{s\rightarrow\infty} g'(s)=0$}, it follows that \m{$g'(s)\leq 0$} for any \m{$s\geq 4$}.
Now, since \m{$g'(s)\leq 0$} for any \m{$s\geq 4$}, and \m{$\lim_{s\rightarrow\infty} g(s)=0$}, it follows that \m{$g(s)\geq 0$} for any \m{$s\geq 4$}.
Therefore, for any \m{$s\in\ZZ^+$}, \m{$s\geq 2$}, we have
\begin{align*}
& \ln\left(1+\frac{2}{s}\right) + (s-1) \ln\left(1-\frac{2}{s^2}\right) = g(s) \geq 0 \\
\Longleftrightarrow
& 1+\frac{2}{s} \geq \left(1-\frac{2}{s^2}\right)^{1-s}
\\
\Longleftrightarrow
& f\left(s,p{=}\frac{2}{s^2}\right)\geq 0,
\end{align*}
as required.
\end{IEEEproof}

\begin{IEEEproof}[Proof of Theorem~\ref{thm:theorem:IndeptProbAccessSymmMaxSpread}]
We will show that if condition~\eqref{eq:MaxCondition} is satisfied, then \m{$\Delta(p,T,k)\geq 0$} for any \m{$k\in\ZZ^+$}.
First, we note that
\begin{align}
\frac{\binom{\floor{kT}}{k-1}}{\binom{\floor{kT}}{k}}
&= \frac{k}{\floor{kT}-k+1} \notag
\\
&= \frac{k}{\floor{k(\floor{T}+\tau)}-k+1},
{\small\hspace{-0.5em}
\begin{array}{l}
\text{where } \tau\triangleq T-\floor{T}\in[0,1)
\end{array}} \notag
\\
&= \frac{k}{k\floor{T}+\floor{k\tau}-k+1} \notag
\\
&\geq \frac{k}{k\floor{T}} \label{eq:MaxBinomFracIneq}
\\
&= \frac{1}{\floor{T}}. \label{eq:MaxBinomFrac}
\end{align}
Inequality~\eqref{eq:MaxBinomFracIneq} follows from the fact that
\[
\floor{k\tau}\leq k\tau<k\;
\Longleftrightarrow\; \floor{k\tau}\leq k-1\;
\Longleftrightarrow\; \floor{k\tau}-k+1\leq 0.
\]
Now, if condition~\eqref{eq:MaxCondition} is satisfied, then we necessarily have \m{$T\geq 2$};
otherwise, \m{$T\in[1,2)$} would imply that
\m{$\floor{T}=1$}, which produces
\m{$(1-p)^{\floor{T}}+2\floor{T}p(1-p)^{\floor{T}-1}-1$}
\m{$=p>0$},
contradicting our assumption.
It follows that
\begin{align}
&\qquad\; (1-p)^{\floor{T}} + 2 \floor{T} p (1-p)^{\floor{T}-1} - 1 \leq 0 \notag
\\
&\Longleftrightarrow \PP{\BB{\floor{T}}{p}=0} + 2\PP{\BB{\floor{T}}{p}=1} - 1 \leq 0 \notag
\\
&\Longleftrightarrow \PP{\BB{\floor{T}}{p}\geq 2} \geq \PP{\BB{\floor{T}}{p}=1} \notag
\\
&\Longleftrightarrow \sum_{j=2}^{\floor{T}} \binom{\floor{T}}{j} p^{j} (1-p)^{\floor{T}-j} \geq \floor{T} p (1-p)^{\floor{T}-1} \notag
\\
&\Longleftrightarrow \sum_{j=2}^{\floor{T}} \frac{1}{\floor{T}} \binom{\floor{T}}{j} \left(\frac{p}{1-p}\right)^{j-1} \geq 1 \label{eq:MaxFloorSum}
\\
&\Longrightarrow \sum_{j=2}^{\ceil{T}} \frac{1}{\floor{T}} \binom{\ceil{T}}{j} \left(\frac{p}{1-p}\right)^{j-1} \geq 1. \label{eq:MaxCeilSum}
\end{align}
Observe that \m{$\alpha_{k,T}\triangleq\floor{(k+1)T}-\floor{kT}$} \m{$\in\left\{\floor{T},\ceil{T}\right\}$}, because \m{$\alpha_{k,T}\in\big(T-1,T+1\big)$} and there are only two integers $\floor{T}$ and $\ceil{T}$, which are possibly nondistinct, in this interval.
It follows from \eqref{eq:MaxFloorSum} and \eqref{eq:MaxCeilSum} that
\begin{align}
\sum_{j=2}^{\alpha_{k,T}} \frac{1}{\floor{T}} \binom{\alpha_{k,T}}{j} \left(\frac{p}{1-p}\right)^{j-1} \geq 1. \label{eq:MaxAlphaSum}
\end{align}
Therefore, we have
\begin{align}
& \sum_{i=1}^{\min(\alpha_{k,T}-1,k)} \sum_{j=i+1}^{\alpha_{k,T}} \frac{\binom{\floor{kT}}{k-i}}{\binom{\floor{kT}}{k}} \binom{\alpha_{k,T}}{j} \left(\frac{p}{1-p}\right)^{-i+j} \notag
\\
&\quad\! \geq \sum_{i=1}^{1} \sum_{j=i+1}^{\alpha_{k,T}} \frac{\binom{\floor{kT}}{k-i}}{\binom{\floor{kT}}{k}} \binom{\alpha_{k,T}}{j}\! \left(\frac{p}{1-p}\right)^{\!\!-i+j} \label{eq:MaxSimplifyMin}
\\
&\quad\! = \sum_{j=2}^{\alpha_{k,T}} \frac{\binom{\floor{kT}}{k-1}}{\binom{\floor{kT}}{k}} \binom{\alpha_{k,T}}{j} \left(\frac{p}{1-p}\right)^{j-1} \notag
\\
&\quad\! \geq \sum_{j=2}^{\alpha_{k,T}} \frac{1}{\floor{T}} \binom{\alpha_{k,T}}{j} \left(\frac{p}{1-p}\right)^{j-1},
\text{ from \eqref{eq:MaxBinomFrac}} \notag
\\
&\quad\! \geq 1,
\text{ from \eqref{eq:MaxAlphaSum}}. \notag
\end{align}
Inequality~\eqref{eq:MaxSimplifyMin} follows from the fact that
\[
\min(\alpha_{k,T}{-}1,k)\;
\geq\; \min(2{-}1,1)\;
=\; 1.
\]
Consequently,
\startcompact{small}
\begin{align*}
& \sum_{i=1}^{\min(\alpha_{k,T}-1,k)} \sum_{j=i+1}^{\alpha_{k,T}} \binom{\floor{kT}}{k-i} \binom{\alpha_{k,T}}{j} \left(\frac{p}{1-p}\right)^{-i+j}
\geq \binom{\floor{kT}}{k}
\\
&{\normalsize
\begin{array}{l}\displaystyle
\Longleftrightarrow \Delta(p,T,k) \geq 0,
\text{ from \eqref{eq:DeltaK}}.
\end{array}}
\end{align*}
\stopcompact{small}
It follows that
\begin{align*}
&\Psm{\floor{T}} \leq
\Psm{\floor{2T}}
\\*
&\hspace{10em} \leq\cdots\leq
\Psm{\floor{\floor{\frac{n}{T}}T}},
\end{align*}
and so we conclude that an optimal $m^*$ is given by either \m{$m=\floor{\floor{\frac{n}{T}}T}$} or \m{$m=n$}.
\end{IEEEproof}

\begin{IEEEproof}[Proof of Corollary~\ref{thm:corollary:IndeptProbAccessSymmMaxSpread}]
If \m{$p\geq\frac{4}{3\floor{T}}$}, then we necessarily have \m{$T\geq 2$};
otherwise, \m{$T\in[1,2)$} would imply that \m{$\floor{T}=1$}, which produces
\m{$p\geq\frac{4}{3\floor{T}}$}
\m{$=\frac{4}{3}$},
contradicting the definition of $p$.
We will show that condition~\eqref{eq:MaxCondition} of Theorem~\ref{thm:theorem:IndeptProbAccessSymmMaxSpread} is satisfied for any \m{$T\geq 2$} and \m{$p\geq\frac{4}{3\floor{T}}$}.
To do this, we define the function
\[
f(p,T) \triangleq (1-p)^{\floor{T}} + 2 \floor{T} p (1-p)^{\floor{T}-1} - 1,
\]
and show that
\m{$f(p,T) \leq f\left(p{=}\frac{4}{3\floor{T}},T\right)\leq 0$}
for any \m{$T\geq 2$} and \m{$p\geq\frac{4}{3\floor{T}}$}.

The partial derivative of \m{$f(p,T)$} wrt $p$ is given by
\[
\frac{\partial}{\partial p} f(p,T)=\floor{T}(1-p)^{\floor{T}-2} \left(1+p-2\floor{T}p\right).
\]
Observe that \m{$f(p,T)$} is decreasing wrt $p$ for any \m{$T\geq 2$} and \m{$p\geq\frac{4}{3\floor{T}}$}, since
\begin{align*}
&\hspace{2em} p\geq\frac{4}{3\floor{T}}
=\frac{1}{\frac{3}{4}\floor{T}}
>\frac{1}{2\floor{T}-1}
\\
&{\small\hspace{-0.5em}
\begin{array}{l}
\Longrightarrow 2\floor{T}p-p>1
\Longleftrightarrow 1+p-2\floor{T}p<0
\Longleftrightarrow \frac{\partial}{\partial p} f(p,T)<0.
\end{array}}
\end{align*}
Now, consider the function
\[
\textstyle
g(T)
\triangleq f\left(p{=}\frac{4}{3\floor{T}},T\right)
= \left(1-\frac{4}{3\floor{T}}\right)^{\floor{T}-1}
\!\!\left(\frac{11}{3}-\frac{4}{3\floor{T}}\right) - 1.
\]
We will proceed to show that \m{$g(T)\leq 0$} for any \m{$T\geq 2$}.
For \m{$T\in[2,3)$}, we have \m{$\floor{T}=2$} and \m{$g(T)=0$}.
To show that \m{$g(T)\leq 0$} for any \m{$T\geq 3$}, we consider the function
\[
h(T)
\triangleq \left(T-1\right) \ln \left(1-\frac{4}{3T}\right) + \ln \left(\frac{11}{3}-\frac{4}{3T}\right), \notag
\]
which has the derivatives
\begin{align*}
h'(T) &= \frac{1}{3T-4} + \frac{11}{11T-4} + \ln \left(1-\frac{4}{3T}\right),
\\
h''(T) &= \frac{16\left(11T^2-24T-16\right)}{T\left(33T^2-56T+16\right)^2}.
\end{align*}
Since \m{$h''(T)>0$} for any \m{$T\geq 3$}, and \m{$\lim_{T\rightarrow\infty} h'(T)=0$}, it follows that \m{$h'(T)\leq 0$} for any \m{$T\geq 3$}.
Now, since \m{$h'(T)\leq 0$} for any \m{$T\geq 3$}, and \m{$h(T{=}3)=$} \m{$\ln\frac{29}{9}-2\ln\frac{9}{5}<0$}, it follows that \m{$h(T)<0$} for any \m{$T\geq 3$}.
Thus, for any \m{$T\geq 3$}, we have
\begin{align*}
& \left(\floor{T}{-}1\right) \ln \left(1-\frac{4}{3\floor{T}}\right) \!+ \ln \left(\frac{11}{3}-\frac{4}{3\floor{T}}\right)
= h(\floor{T}) < 0
\\
&\Longleftrightarrow \ln\left\{ \left(1-\frac{4}{3\floor{T}}\right)^{\floor{T}-1} \left(\frac{11}{3}-\frac{4}{3\floor{T}}\right) \right\} < 0
\\
&\Longleftrightarrow
\left(1-\frac{4}{3\floor{T}}\right)^{\!\!\floor{T}-1} \left(\frac{11}{3}-\frac{4}{3\floor{T}}\right) < 1
\Longleftrightarrow
g(T) < 0.
\end{align*}
Combining these results, we obtain
\[
f(p,T)
\leq f\left(p{=}\frac{4}{3\floor{T}},T\right)
= g(T) \leq 0
\]
for any \m{$T\geq 2$} and \m{$p\geq\frac{4}{3\floor{T}}$}, as required.
\end{IEEEproof}

\begin{IEEEproof}[Proof of Lemma~\ref{thm:lemma:IndeptProbAccessSymmMinSpread1}]
Suppose that \m{$T>1$}.
We will show that if condition~\eqref{eq:MinCondition1} or condition~\eqref{eq:MinCondition2} is satisfied, then \m{$\Delta(p,T,k)\leq 0$} for any \m{$k\in\ZZ^+$}.
First, we note that for any $i\in$ \m{$\{1,\ldots,k\}$},
\begin{align}
\frac{\binom{\floor{kT}}{k-i}}{\binom{\floor{kT}}{k}}
&= \frac{\overbrace{(k)(k-1)\!\cdots\!(k-i+1)}^{i \text{ terms}}}{\underbrace{(\floor{kT}-k+i)\!\cdots\!(\floor{kT}-k+2)(\floor{kT}-k+1)}_{i \text{ terms}}} \notag
\\
&\leq \left(\frac{k}{\floor{kT}-k+1}\right)^{i} \notag
\\
&\leq \left(\frac{k}{kT-1-k+1}\right)^{i} \notag
\\
&= \left(\frac{1}{T-1}\right)^{i}. \label{eq:MinBinomFrac}
\end{align}
Now, if condition~\eqref{eq:MinCondition1} is satisfied, then
\begin{align*}
&\;\;\; \sum_{i=1}^{\ceil{T}-1} \sum_{j=i+1}^{\ceil{T}} \left(\frac{1}{T-1}\right)^{i} \binom{\ceil{T}}{j} \left(\frac{p}{1-p}\right)^{-i+j}
\\
&= \sum_{i=1}^{T-1} \sum_{j=i+1}^{T} \left(\frac{1}{T-1}\right)^{i} \binom{T}{j} \left(\frac{\frac{1}{T}}{1-\frac{1}{T}}\right)^{-i+j}
\\
&= \sum_{i=1}^{T-1} \sum_{j=i+1}^{T} \binom{T}{j} \left(\frac{1}{T-1}\right)^{j}
\\
&= \sum_{\ell=2}^{T} (\ell-1) \binom{T}{\ell} \left(\frac{1}{T-1}\right)^{\ell} \,
= \,
1.
\end{align*}
On the other hand, if condition~\eqref{eq:MinCondition2} is satisfied, then
\begin{align*}
&\;\;\;\; \sum_{i=1}^{\ceil{T}-1} \sum_{j=i+1}^{\ceil{T}} \left(\frac{1}{T-1}\right)^{i} \binom{\ceil{T}}{j} \left(\frac{p}{1-p}\right)^{-i+j}
\\
&= \sum_{i=1}^{\ceil{T}-1} \sum_{j=i+1}^{\ceil{T}} \binom{\ceil{T}}{j} \left(\frac{1-p}{p(T-1)}\right)^{i}  \left(\frac{p}{1-p}\right)^{j}
\\
&= \sum_{\ell=2}^{\ceil{T}} \left(\sum_{r=1}^{\ell-1} \left(\frac{1-p}{p(T-1)}\right)^{r}\right) \binom{\ceil{T}}{\ell} \left(\frac{p}{1-p}\right)^{\ell}
\\
&= 1 -\frac{T\left(\frac{1}{T}\left(1-\frac{1}{T}\right)^{\ceil{T}-1}-p(1-p)^{\ceil{T}-1}\right)}{(1-pT)\left(1-\frac{1}{T}\right)^{\ceil{T}-1}(1-p)^{\ceil{T}-1}} \;
\leq \;
1.
\end{align*}
Thus, if either condition is satisfied, we have
\begin{align}
&\qquad\; \sum_{i=1}^{\ceil{T}-1} \sum_{j=i+1}^{\ceil{T}} \left(\frac{1}{T-1}\right)^{\!\!i} \binom{\ceil{T}}{j} \left(\frac{p}{1-p}\right)^{\!\!-i+j} \!\!\leq 1 \label{eq:MinCeilSum}
\\
&\Longrightarrow \sum_{i=1}^{\floor{T}-1} \sum_{j=i+1}^{\floor{T}} \left(\frac{1}{T-1}\right)^{\!\!i} \binom{\floor{T}}{j} \left(\frac{p}{1-p}\right)^{\!\!-i+j} \!\!\leq 1. \label{eq:MinFloorSum}
\end{align}
As in the proof of Theorem~\ref{thm:theorem:IndeptProbAccessSymmMaxSpread}, we note that \m{$\alpha_{k,T}\triangleq$} \m{$\floor{(k+1)T}-\floor{kT}$} \m{$\in\left\{\floor{T},\ceil{T} \right\}$}.
It follows from \eqref{eq:MinCeilSum} and \eqref{eq:MinFloorSum} that
\begin{align}
\sum_{i=1}^{\alpha_{k,T}-1} \sum_{j=i+1}^{\alpha_{k,T}} \left(\frac{1}{T-1}\right)^{i} \binom{\alpha_{k,T}}{j} \left(\frac{p}{1-p}\right)^{-i+j} \leq 1. \label{eq:MinAlphaSum}
\end{align}
Therefore, we have
\begin{align*}
&\sum_{i=1}^{\min(\alpha_{k,T}-1,k)} \sum_{j=i+1}^{\alpha_{k,T}} \frac{\binom{\floor{kT}}{k-i}}{\binom{\floor{kT}}{k}} \binom{\alpha_{k,T}}{j} \left(\frac{p}{1-p}\right)^{-i+j}
\\
&\leq \hspace{-1.2em} \sum_{i=1}^{\min(\alpha_{k,T}-1,k)} \sum_{j=i+1}^{\alpha_{k,T}} \!\! \left(\frac{1}{T-1}\right)^{\!\!i} \!\! \binom{\alpha_{k,T}}{j} \!\! \left(\frac{p}{1-p}\right)^{\!\!-i+j}\!\!\!,
{\small
\begin{array}{l}
\text{from \eqref{eq:MinBinomFrac}}
\end{array}}
\\
&\leq \sum_{i=1}^{\alpha_{k,T}-1} \sum_{j=i+1}^{\alpha_{k,T}} \!\! \left(\frac{1}{T-1}\right)^{\!\!i} \!\! \binom{\alpha_{k,T}}{j} \!\! \left(\frac{p}{1-p}\right)^{\!\!-i+j} \notag
\\
&\leq 1,
\text{ from \eqref{eq:MinAlphaSum}}.
\end{align*}
Consequently,
\startcompact{small}
\begin{align*}
& \sum_{i=1}^{\min(\alpha_{k,T}-1,k)} \sum_{j=i+1}^{\alpha_{k,T}} \binom{\floor{kT}}{k-i} \binom{\alpha_{k,T}}{j} \left(\frac{p}{1-p}\right)^{-i+j}
\leq \binom{\floor{kT}}{k}
\\
&{\normalsize
\begin{array}{l}\displaystyle
\Longleftrightarrow \Delta(p,T,k) \leq 0,
\text{ from \eqref{eq:DeltaK}}.
\end{array}}
\end{align*}
\stopcompact{small}
It follows that
\begin{align*}
&\Psm{\floor{T}} \geq
\Psm{\floor{2T}}
\\*
&\hspace{10em} \geq
\Psm{\floor{3T}} \geq\cdots,
\end{align*}
and since
\begin{align}
\Psm{n}
\begin{cases}
= \Psm{\floor{\floor{\frac{n}{T}}T}}
& \hspace{-0.8em}\text{if } \frac{n}{T}\in\ZZ^+\!,
\\
\leq \Psm{\floor{\left(\floor{\frac{n}{T}}+1\right)T}}
& \hspace{-0.8em}\text{otherwise,}
\end{cases} \notag
\end{align}
we conclude that an optimal $m^*$ is given by \m{$m=\floor{T}$}.
\end{IEEEproof}

\begin{IEEEproof}[Proof of Lemma~\ref{thm:lemma:IndeptProbAccessSymmMinSpread2}]
Since \xxsm{\floor{T}} is indeed optimal for \emph{any} $p$ when \m{$T=1$},
we need only consider the case of \m{$T>1$}.
We will show that either condition~\eqref{eq:MinCondition1} or condition~\eqref{eq:MinCondition2} of Lemma~\ref{thm:lemma:IndeptProbAccessSymmMinSpread1} is satisfied for any \m{$T>1$} and \m{$p\leq\frac{2}{\ceil{T}}-\frac{1}{T}$}.
We do this in two steps:
First, we define the function
\[
f(p,T) \triangleq \frac{p(1-p)^{\ceil{T}-1}}{\frac{1}{T}\left(1-\frac{1}{T}\right)^{\ceil{T}-1}} - 1,
\]
and show that
\m{$f(p,T) \leq f\left(p{=}\frac{2}{\ceil{T}}{-}\frac{1}{T},T\right) \leq 0$} for any \m{$T>1$} and \m{$p\leq\frac{2}{\ceil{T}}-\frac{1}{T}$}.
Second, we apply the appropriate condition from Lemma~\ref{thm:lemma:IndeptProbAccessSymmMinSpread1} for each pair of $T$ and $p$.

The partial derivative of $f(p,T)$ wrt $p$ is given by
\[
\frac{\partial}{\partial p} f(p,T) = \frac{\left(1-p\ceil{T}\right)(1-p)^{\ceil{T}-2}}{\frac{1}{T}\left(1-\frac{1}{T}\right)^{\ceil{T}-1}}. \]
Observe that \m{$f(p,T)$} is nondecreasing wrt $p$ for any \m{$T>1$} and \m{$p\leq\frac{2}{\ceil{T}}-\frac{1}{T}$}, since
\begin{align*}
&\hspace{2em} p\leq\frac{2}{\ceil{T}}-\frac{1}{T}
\leq\frac{2}{\ceil{T}}-\frac{1}{\ceil{T}}
=\frac{1}{\ceil{T}}
\\
&\Longrightarrow p\ceil{T}\leq 1
\Longleftrightarrow 1-p\ceil{T}\geq 0
\Longleftrightarrow \frac{\partial}{\partial p} f(p,T)\geq 0.
\end{align*}
Now, consider the function
\[
\textstyle
g(T)
\triangleq
f\left(p{=}\frac{2}{\ceil{T}}{-}\frac{1}{T},T\right)
=
\frac{\left(\frac{2}{\ceil{T}}-\frac{1}{T}\right)
\left(1-\frac{2}{\ceil{T}}+\frac{1}{T}\right)^{\ceil{T}-1}}
{\frac{1}{T}\left(1-\frac{1}{T}\right)^{\ceil{T}-1}} - 1.
\]
We will proceed to show that \m{$g(T)\leq 0$} for any \m{$T>1$} by reparameterizing \m{$g(T)$} as \m{$h(c,\tau)$}, where \m{$c\triangleq\ceil{T}$} and \m{$\tau\triangleq\ceil{T}-T$}:
\[
h(c,\tau)
\triangleq
g\left(T{=}c{-}\tau\right)
=
\frac{\left(\frac{2}{c}-\frac{1}{c-\tau}\right)\!\left(1-\frac{2}{c}+\frac{1}{c-\tau}\right)^{\!c-1}}{\frac{1}{c-\tau}\left(1-\frac{1}{c-\tau}\right)^{\!c-1}} - 1.
\]
The partial derivative of \m{$h(c,\tau)$} wrt $\tau$ is given by
\[
\frac{\partial}{\partial\tau} h(c,\tau)
=
-\frac{2\tau^{2}(c-2) \left(1-\frac{2}{c}+\frac{1}{c-\tau}\right)^{c}}{\big(c(c-1-\tau)+2\tau\big)^{2}\left(1-\frac{1}{c-\tau}\right)^{c}}.
\]
Since \m{$\frac{\partial}{\partial\tau} h(c,\tau) \leq 0$} for any
\m{$c\in\ZZ^+$}, \m{$c\geq 2$}, and \m{$\tau\in[0,1)$}, it follows that for any \m{$T>1$}, we have
\begin{align*}
g(T)
&= h\left(c{=}\ceil{T},\tau{=}\ceil{T}{-}T\right)
\\
&\leq h\left(c{=}\ceil{T},\tau{=}0\right)
\\
&= \frac{\left(\frac{2}{\ceil{T}}-\frac{1}{\ceil{T}}\right)\left(1-\frac{2}{\ceil{T}}+\frac{1}{\ceil{T}}\right)^{\ceil{T}-1}}{\frac{1}{\ceil{T}}\left(1-\frac{1}{\ceil{T}}\right)^{\ceil{T}-1}} - 1
= 0.
\end{align*}
Combining these results, we obtain
\[
f(p,T)
\leq f\left(p{=}\frac{2}{\ceil{T}}{-}\frac{1}{T},T\right)
= g(T)
\leq 0
\]
for any \m{$T>1$} and \m{$p\leq\frac{2}{\ceil{T}}-\frac{1}{T}$}, which implies
\[
p\left(1-p\right)^{\ceil{T}-1}
\leq \frac{1}{T}\left(1-\frac{1}{T}\right)^{\ceil{T}-1}.
\]
Finally, we apply the appropriate condition from Lemma~\ref{thm:lemma:IndeptProbAccessSymmMinSpread1} for each pair of $T$ and $p$.
For \m{$T\in\ZZ^+$}, \m{$T>1$}, we have \m{$\frac{2}{\ceil{T}}-\frac{1}{T}=\frac{1}{T}$}:
we use condition~\eqref{eq:MinCondition1} for \m{$p=\frac{1}{T}$}, and condition~\eqref{eq:MinCondition2} for \m{$p<\frac{1}{T}$}.
For \m{$T\notin\ZZ^+$}, \m{$T>1$}, we have \m{$\frac{2}{\ceil{T}}-\frac{1}{T}<\frac{1}{T}$}:
we use condition~\eqref{eq:MinCondition2} for \m{$p<\frac{1}{T}$}.
\end{IEEEproof}

\begin{IEEEproof}[Proof of Theorem~\ref{thm:theorem:IndeptProbAccessSymmMinSpread}]
Since \xxsm{\floor{T}} is indeed optimal for \emph{any} $p$ when \m{$T=1$},
we need only consider the case of \m{$T>1$}.
We will show that \xxsm{\floor{T}} is an optimal symmetric allocation for any \m{$T>1$} and \m{$p\leq\frac{1}{\ceil{T}}$}.
We do this by considering subintervals of $T$ over which $\ceil{T}$ is constant.

Let $T$ be confined to the unit interval \m{$(c,c+1]$}, where \m{$c\in\ZZ^+$}.
According to Lemma~\ref{thm:lemma:IndeptProbAccessSymmMinSpread2}, \xxsm{\floor{T}} is optimal for any
\m{$p\in\left(0,\frac{2}{c+1}-\frac{1}{T}\right]$} and \m{$T\in(c,c+1]$},
or equivalently, for any
\[
p\in\left(0,\frac{1}{c+1}\right]
\text{~~and~~}
T\in\left[\frac{1}{\frac{2}{c+1}-p},c+1\right] \cap (c,c+1].
\]
This is just the area below a ``peak'' in Fig.~\ref{fig:IndeptProbAccessOptSymmRegionPlot}, expressed in terms of different independent variables.
For each
\m{$p\in\left(0,\frac{1}{c+1}\right)$}, we can always find a $T_0$ such that
\[
T_0\in\left[\frac{1}{\frac{2}{c+1}-p},c+1\right) \cap (c,c+1).
\]
For example, we can pick \m{$T_0=c+1-\delta$}, where
\[
\delta
\triangleq\frac{1}{2}\left(c+1-\max\left(c,\frac{1}{\frac{2}{c+1}-p}\right)\right)
\in(0,1).
\]
Now, we make the crucial observation that if \xxsm{\floor{T}} is an optimal symmetric allocation for \m{$T=T_0$}, then \xxsm{\floor{T}} is also an optimal symmetric allocation for any \m{$T\in\big[\floor{T_0},T_0\big]$}.
This claim can be proven by contradiction:
the recovery probability for \xxsm{\floor{T}} is given by
\[
\Psm{\floor{T}}=\PP{\BB{\floor{T}}{p}\geq 1}
\]
which remains constant for all
\m{$T\in\big[\floor{T_0},T_0\big]$}, and a symmetric allocation that performs strictly better than
\xxsm{\floor{T}} for some \m{$T\in\big[\floor{T_0},T_0\big]$} would therefore also outperform \xxsm{\floor{T}} for \m{$T=T_0$}.
Since \xxsm{\floor{T}} is indeed optimal for our choice of $T_0$, it follows then that \xxsm{\floor{T}} is also optimal for any
\[
p\in\left(0,\frac{1}{c+1}\right)
\text{~~and~~}
T\in(c,c+1].
\]
By applying this result for each \m{$c\in\ZZ^+$}, we reach the conclusion that \xxsm{\floor{T}} is an optimal symmetric allocation for any \m{$T>1$} and \m{$p<\frac{1}{\ceil{T}}$}.

Finally, to extend the optimality of \xxsm{\floor{T}} to \m{$p=\frac{1}{\ceil{T}}$}, we note that the recovery probability
\m{$\Ps(p,T,m)$} $\triangleq$ \m{$\PP{\BB{m}{p}\geq\ceil{\frac{m}{T}}}$}
is a polynomial in $p$ and is therefore continuous at \m{$p=\frac{1}{\ceil{T}}$}.
Since \xxsm{\floor{T}} is optimal as \m{$p\rightarrow\frac{1}{\ceil{T}}^-$}, it remains optimal at \m{$p=\frac{1}{\ceil{T}}$}.
\end{IEEEproof}

\begin{IEEEproof}[Proof of Proposition~\ref{thm:proposition:RandFixedSizeSubsetPComplete}]
Consider an allocation $\xxn$ where each $x_i$ is a nonnegative rational number.
The problem of computing the recovery probability for this allocation and a given subset size $r$ is equivalent to the counting version of the following decision problem (which happens to be polynomial-time solvable):

\begin{thm:definition}
\textsc{Largest $r$-Subset Sum} (\textsc{LRSS})
\\
\textit{Instance}:
Finite $n$-vector \m{$\left(a_1,\ldots,a_n\right)$} with \m{$a_i\in\ZZ^+_0$}, file size \m{$d\in\ZZ^+$}, and subset size \m{$r\in\ZZ^+$}, where all $a_i$ and $d$ can be written as decimal numbers of length at most $\ell$.
\\
\textit{Question}:
Is there an $r$-subset \m{$\rr\subseteq\nset$} that satisfies
\m{$\sum_{i\in\rr} a_i\geq d$}?
\end{thm:definition}

Note that the allocation and file size have been scaled so that the problem parameters are all integers.
To show that the counting problem \textsc{\#LRSS} is \#P-complete, we essentially apply the proof of Proposition~\ref{thm:proposition:IndeptProbAccessPHard}, substituting \text{\#LSS} with \text{\#LRSS}, and stipulating that the subset size \m{$r=m+k$} in the Turing reduction.
\end{IEEEproof}

\begin{IEEEproof}[Proof of Lemma~\ref{thm:lemma:RandFixedSizeSubsetCover}]
Summing up the $c$ inequalities of \eqref{eq:inequalitiesm} produces
\[
\sum_{j=1}^{c} \sum_{i\in\rr_j} x_i \geq c.
\]
The terms on the left-hand side can be regrouped to obtain
\[
\sum_{i\in S} \sum_{j=1}^{c} \II{i\in\rr_j}\, x_i \geq c.
\]
Substituting \eqref{eq:variablecount} into the above inequality yields
\[
\sum_{i\in S} b\, x_i \geq c,
\]
as required.
\end{IEEEproof}

\begin{IEEEproof}[Proof of Lemma~\ref{thm:lemma:RandFixedSizeSubsetProb1}]
Let $\RRR$ be the collection of all $\binom{n}{r}$ possible \m{$r$-subsets} of $\nset$.
If \m{$\Ps=1$}, then any feasible allocation must satisfy
\[
\sum_{i\in\rr} x_i \geq 1 \quad
\forall\;\;
\rr\in\RRR.
\]
Observe that each element in $\nset$ appears the same number of times among the \m{$r$-subsets} in $\RRR$.
Specifically, the number of \m{$r$-subsets} that contain element \m{$i\in\nset$} is just the number of ways of choosing the other \m{$(r-1)$} elements of the \m{$r$-subset} from the remaining \m{$(n-1)$} elements of $\nset$, i.e.,
\[
\sum_{\rr\in\RRR} \II{i\in\rr} = \binom{n-1}{r-1} \quad \forall\;\; i\in\nset.
\]
Applying Lemma~\ref{thm:lemma:RandFixedSizeSubsetCover} with \m{$S=\nset$}, \m{$c=\binom{n}{r}$}, and \m{$b=\binom{n-1}{r-1}$} therefore produces
\[
\sum_{i=1}^{n} x_i \geq \frac{\binom{n}{r}}{\binom{n-1}{r-1}} = \frac{n}{r}
\]
for any feasible allocation.
Now, $\xxr$ is a feasible allocation since it has a recovery probability of exactly $1$;
because it uses the minimum possible total amount of storage $\frac{n}{r}$, this allocation is also optimal.
\end{IEEEproof}

\begin{IEEEproof}[Proof of Theorem~\ref{thm:theorem:RandFixedSizeSubsetIffDiv}]
Suppose that $n$ is a multiple of $r$;
let positive integer $\alpha$ be defined such that
\m{$n=\alpha r$}.

We will first prove that \m{$\Ps>1-\frac{r}{n}$} is a sufficient condition for the optimality of $\xxr$ by showing that if the constraint
\begin{align}
\sum_{i\in\rr} x_i \geq 1 \label{eq:RecovConstraint}
\end{align}
is satisfied for more than
\m{$\left(1-\frac{r}{n}\right) \binom{n}{r}$}
distinct \m{$r$-subsets} \m{$\rr\subseteq\nset$}, then the allocation $\xxr$ minimizes the required budget $T$.
Our approach is motivated by the observation of Lemma~\ref{thm:lemma:RandFixedSizeSubsetCover}.
We begin by constructing a collection of \m{$r$-subsets} such that if constraint~\eqref{eq:RecovConstraint} is satisfied for the \m{$r$-subsets} in this collection, then
\m{$\sum_{i=1}^{n} x_i \geq\frac{n}{r}$}.
We then demonstrate that such a collection of \m{$r$-subsets} can be found among \emph{any} collection of more than \m{$\left(1-\frac{r}{n}\right) \binom{n}{r}$} distinct \m{$r$-subsets}.

Let
\[
Q \triangleq \left(\vv_1,\ldots,\vv_{\alpha}\right)
\]
be an ordered partition of $\nset$ that comprises $\alpha$ parts, where
\m{$|\vv_j|=r$}, \m{$j=1,\ldots,\alpha$}.
For a given ordered partition $Q$, we specify a collection of $\alpha$ distinct \m{$r$-subsets}
\begin{align*}
\RRR_Q
&\triangleq \{\rr_1,\ldots,\rr_{\alpha}\},
\\*
\text{where }
\rr_j
&\triangleq \vv_j, \quad
j=1,\ldots,\alpha.
\end{align*}
\begin{figure}
\centering
\includegraphics[clip=true, trim=2mm 5mm 2mm 2mm, width=0.27\textwidth]{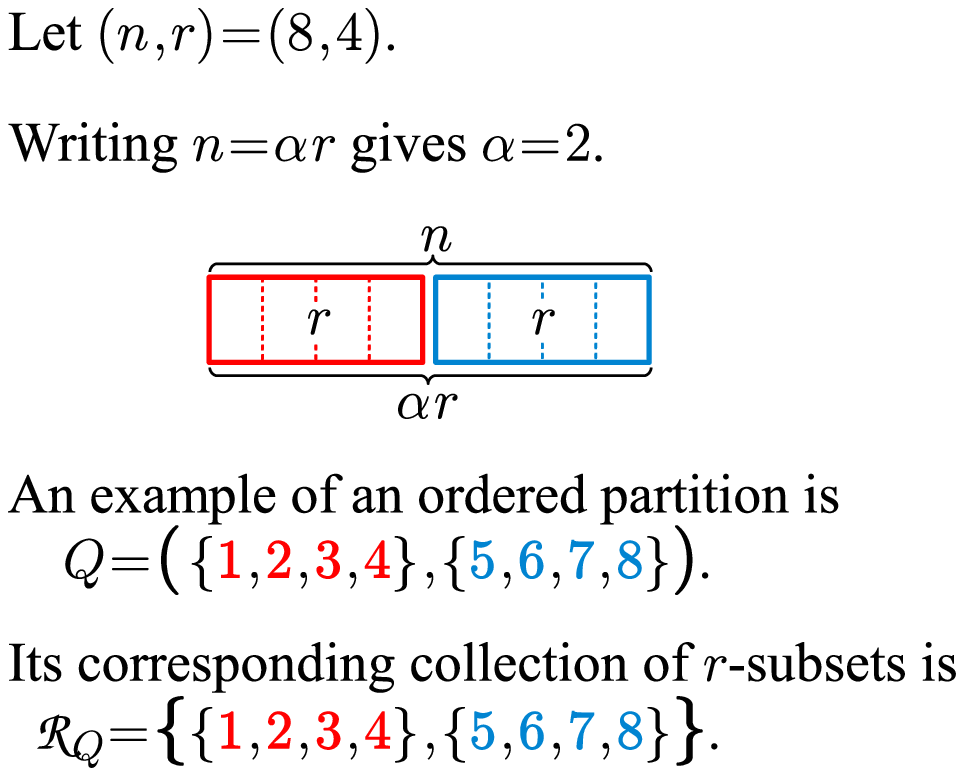}
\caption{Example for the construction of the ordered partition $Q$ and its corresponding collection of \m{$r$-subsets} $\RRR_Q$, in the proof of Theorem~\ref{thm:theorem:RandFixedSizeSubsetIffDiv} (when $n$ is a multiple of $r$).}
\label{fig:RandFixedSizeSubsetProofDiv}
\end{figure}%
Fig.~\ref{fig:RandFixedSizeSubsetProofDiv} provides an example of how $Q$ and $\RRR_Q$ are constructed.
Let $A$ be the total number of possible ordered partitions $Q$.
By counting the number of ways of picking $\vv_j$, we have
\[
A =
\underbrace{%
\binom{\alpha r}{r}
\binom{(\alpha-1)r}{r}
\binom{(\alpha-2)r}{r}\cdots
\binom{r}{r}}%
_{\alpha \text{ terms}}
=\frac{(\alpha r)!}{(r!)^{\alpha}}.
\]
Let $B$ be the number of ordered partitions $Q$ for which \m{$\rr\in\RRR_Q$}, for a given \m{$r$-subset} \m{$\rr\subseteq\nset$}.
By counting the number of ways of picking $\vv_j$, subject to the requirement that \m{$\rr\in\RRR_Q$}, we have
\[
B =
\alpha \underbrace{%
\binom{(\alpha-1)r}{r}
\binom{(\alpha-2)r}{r}\cdots
\binom{r}{r}}%
_{(\alpha-1) \text{ terms}}
=\frac{\alpha\big((\alpha-1) r\big)!}{(r!)^{\alpha-1}}.
\]

We claim that for any given ordered partition $Q$, if
\[
\sum_{i \in \rr} x_i \geq 1 \quad \forall\;\; \rr\in \RRR_Q,
\]
then \m{$\sum_{i=1}^{n} x_i \geq\frac{n}{r}$}.
To see this, observe that each element \m{$i\in\nset$} appears in exactly one of the $\alpha$ \m{$r$-subsets} of $\RRR_Q$, i.e.,
\[
\sum_{\rr\in \RRR_Q} \II{i\in\rr} = 1 \quad \forall\;\; i\in\nset.
\]
Applying Lemma~\ref{thm:lemma:RandFixedSizeSubsetCover} with \m{$S=\nset$}, \m{$c=\alpha$}, and \m{$b=1$} therefore produces
\m{$\sum_{i=1}^{n} x_i \geq\frac{\alpha}{1}=\frac{n}{r}$}.

Let $\RRR$ be the collection of all $\binom{n}{r}$ possible \m{$r$-subsets} of $\nset$.
Observe that all $A$ collections $\RRR_Q$ can be found in $\RRR$, i.e.,
\[
\RRR_{Q_1}\subseteq\RRR,\quad
\RRR_{Q_2}\subseteq\RRR,\quad
\ldots,\quad
\RRR_{Q_A}\subseteq\RRR.
\]
With each removal of an \m{$r$-subset} from $\RRR$, we reduce the number of collections $\RRR_Q$ that can be found among the remaining \m{$r$-subsets} by at most $B$.
It follows that the minimum number of \m{$r$-subsets} that need to be removed from $\RRR$ so that no collections $\RRR_Q$ remain is at least $\ceil{\frac{A}{B}}$, where
\[
\frac{A}{B}
= \frac{(\alpha r)!}{\alpha\, r! \big((\alpha-1)r\big)!}
= \frac{r}{n} \binom{n}{r}.
\]
Thus, if fewer than
\m{$\frac{A}{B}=\frac{r}{n}\binom{n}{r}$}
\m{$r$-subsets} are removed from $\RRR$, then at least one collection $\RRR_Q$ would remain;
equivalently, some collection $\RRR_Q$ can be found among \emph{any} collection of more than \m{$\left(1-\frac{r}{n}\right)\binom{n}{r}$} distinct \m{$r$-subsets}.

We have therefore shown that if
\m{$\Ps>1-\frac{r}{n}$},
then any feasible allocation must satisfy
\m{$\sum_{i=1}^{n} x_i\geq\frac{n}{r}$}.
Now, $\xxr$ is a feasible allocation since it has a recovery probability of exactly $1$;
because it uses the minimum possible total amount of storage $\frac{n}{r}$, this allocation is also optimal.

We proceed to prove that
\m{$\Ps>1-\frac{r}{n}$}
is also a necessary condition for the optimality of $\xxr$ by demonstrating that this allocation is suboptimal for any
\m{$\Ps\leq 1-\frac{r}{n}$}.

For \m{$r<n$}, the allocation
\m{$\left(0,\frac{1}{r},\ldots,\frac{1}{r}\right)$}
has a recovery probability of
\m{$\binom{n-1}{r}\big/\binom{n}{r}$} \m{$=1-\frac{r}{n}$} and is therefore a feasible allocation for any
\m{$\Ps\leq 1-\frac{r}{n}$}.
Since this allocation uses a smaller total amount of storage
\m{$\frac{n-1}{r}<\frac{n}{r}$}, it is a strictly better allocation than $\xxr$ for any
\m{$\Ps\leq 1-\frac{r}{n}$}.

For the trivial case \m{$r=n$}, we have \m{$1-\frac{r}{n}=0$}.
The empty allocation \m{$\left(0,\ldots,0\right)$} is clearly optimal for any \m{$\Ps\leq 0$}.
\end{IEEEproof}

\begin{IEEEproof}[Proof of Theorem~\ref{thm:theorem:RandFixedSizeSubsetNotDiv}]
Suppose that $n$ is not a multiple of $r$;
let integers $\alpha$ and $r'$ be as defined in the theorem.
For brevity, we additionally define positive integers $d$, $m$, and $m'$ such that
\[
d=\gcd(r,r'),\quad
r=m\,d,\quad
r'=m'\,d.
\]
We can therefore write \m{$n=(\alpha\,m+m')d$}.

We will prove that
\[
\Ps > 1-\frac{d}{\alpha\,d + m'\,d}
= 1-\frac{1}{\alpha+m'}
\]
is a sufficient condition for the optimality of $\xxr$ by showing that if the constraint
\[
\sum_{i\in\rr} x_i \geq 1
\]
is satisfied for more than
\m{$\left(1-\frac{1}{\alpha+m'}\right) \binom{n}{r}$}
distinct \m{$r$-subsets} \m{$\rr\subseteq\nset$}, then the allocation $\xxr$ minimizes the required budget $T$.
We apply the proof technique of Theorem~\ref{thm:theorem:RandFixedSizeSubsetIffDiv}, but modify the construction of the ordered partition $Q$ and its corresponding collection of \m{$r$-subsets} $\RRR_Q$ to take into account the indivisibility of $n$ by $r$.

For the moment, we will proceed with the assumption that \m{$\alpha\geq 1$}.
Let
\[
Q \triangleq (\uu_1,\ldots,\uu_{m'},\vv_1,\ldots,\vv_{\alpha})
\]
be an ordered partition of $\nset$ that comprises \m{$(m'+\alpha)$} parts, where
\begin{align*}
|\uu_j| &= d, &j&=1,\ldots,m',
\\
|\vv_j| &= r = m\,d, &j&=1,\ldots,\alpha.
\end{align*}
For a given ordered partition $Q$, we specify a collection of \m{$(m'+\alpha)$} distinct \m{$r$-subsets}
\begin{align*}
\RRR_Q
&\triangleq \{\rr_1,\ldots,\rr_{m'},\rr_{m'+1},\ldots,\rr_{m'+\alpha}\},
\\
\text{where }
\rr_j
&\triangleq
\begin{cases}
\displaystyle\bigcup_{\ell=0}^{m-1} \uu_{j+\ell} & \text{if } j=1,\ldots,m',\\
\vv_{j-m'} & \text{if } j=m'+1,\ldots,m'+\alpha,
\end{cases}
\\*
\text{and }
\uu_j
&\triangleq
\uu_{j-m'}
\text{ if }
j>m'.
\end{align*}
\begin{figure}
\centering
\includegraphics[clip=true, trim=12mm 5mm 2mm 2mm, width=0.35\textwidth]{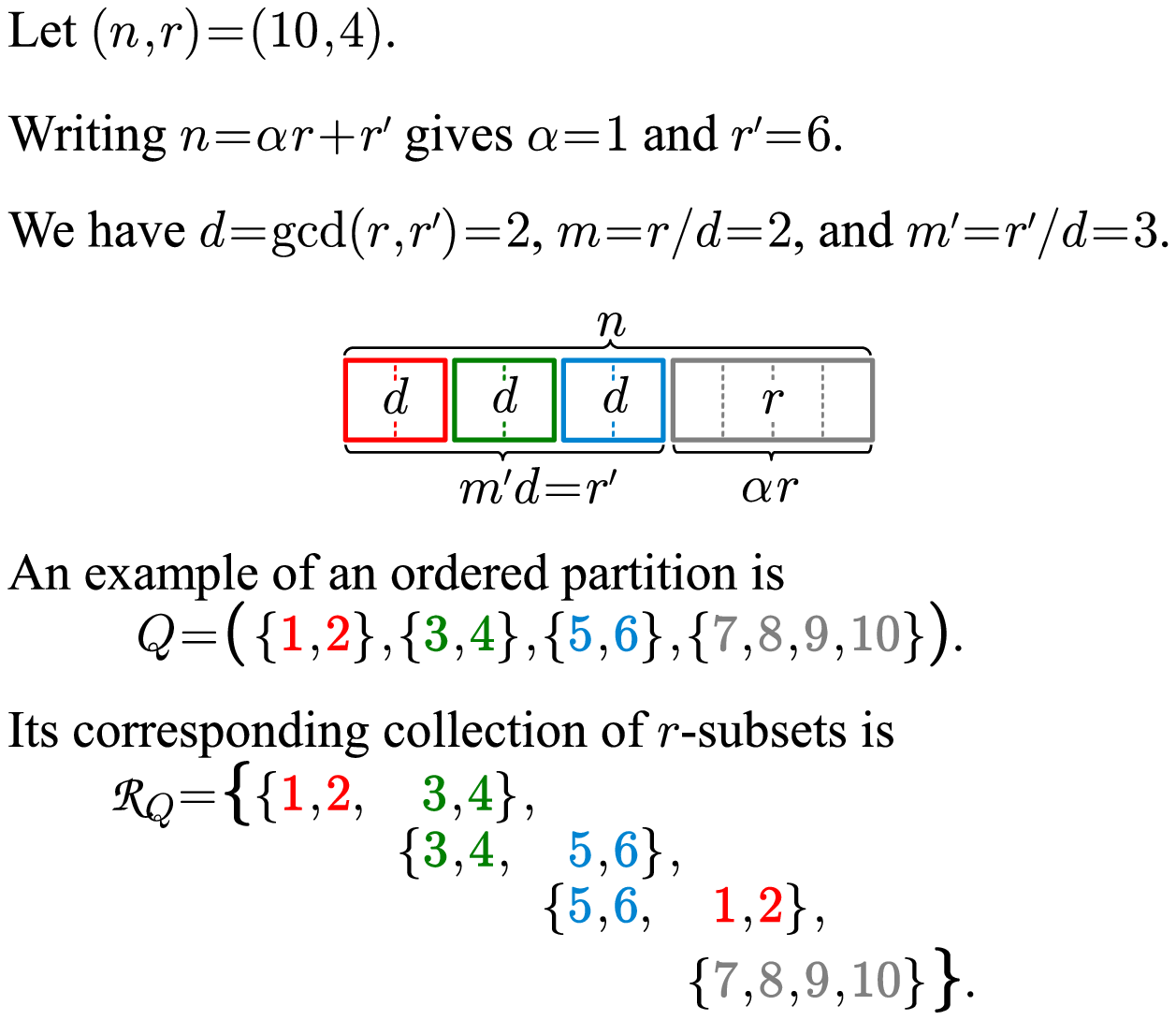}
\caption{Example for the construction of the ordered partition $Q$ and its corresponding collection of \m{$r$-subsets} $\RRR_Q$, in the proof of Theorem~\ref{thm:theorem:RandFixedSizeSubsetNotDiv} (when $n$ is not a multiple of $r$).}
\label{fig:RandFixedSizeSubsetProofNotDiv}
\end{figure}%
Fig.~\ref{fig:RandFixedSizeSubsetProofNotDiv} provides an example of how $Q$ and $\RRR_Q$ are constructed.
Let $A$ be the total number of possible ordered partitions $Q$.
By counting the number of ways of picking $\uu_j$ and $\vv_j$, we have
\begin{align*}
& A =
\underbrace{%
\binom{(\alpha m{+}m')d}{d}
\binom{(\alpha m{+}m'{-}1)d}{d}\cdots
\binom{(\alpha m{+}1)d}{d}}%
_{m' \text{ terms}}\cdot
\\*
&\underbrace{%
\binom{\alpha md}{md}
\binom{(\alpha{-}1)md}{md}\cdots
\binom{md}{md}}%
_{\alpha \text{ terms}}
= \frac{\big((\alpha m+m')d\big)!}{(d!)^{m'} \big((md)!\big)^{\alpha}}.
\end{align*}
Let $B$ be the number of ordered partitions $Q$ for which \m{$\rr\in\RRR_Q$}, for a given \m{$r$-subset} \m{$\rr\subseteq\nset$}.
By counting the number of ways of picking $\uu_j$ and $\vv_j$, subject to the requirement that \m{$\rr\in\RRR_Q$}, we have
\startcompact{footnotesize}
\begin{align*}
& B \!=\!
\underbrace{%
\binom{\big((\alpha{-}1)m{+}m'\big)d}{d}\!\!
\binom{\big((\alpha{-}1)m{+}m'{-}1\big)d}{d}\!\!\cdots\!\!
\binom{\big((\alpha{-}1)m{+}1\big)d}{d}}%
_{m' \text{ terms}}\cdot
\\*
&\; \alpha\underbrace{%
\binom{(\alpha{-}1)md}{md}
\binom{(\alpha{-}2)md}{md}\cdots
\binom{md}{md}}%
_{(\alpha-1) \text{ terms}}
\\
& + \quad m' \underbrace{%
\binom{md}{d}
\binom{(m{-}1)d}{d}\cdots
\binom{d}{d}}%
_{m \text{ terms}}\cdot
\\*
& \underbrace{%
\binom{\big((\alpha{-}1)m{+}m'\big)d}{d}
\binom{\big((\alpha{-}1)m{+}m'{-}1\big)d}{d}\cdots
\binom{(\alpha m{+}1)d}{d}%
}_{(m'-m) \text{ terms}}\cdot
\\*
& \underbrace{%
\binom{\alpha md}{md}
\binom{(\alpha{-}1)md}{md}\cdots
\binom{md}{md}}%
_{\alpha \text{ terms}}
\\
&{\normalsize
\begin{array}{l}\displaystyle
= \alpha\frac{\big(\big((\alpha-1)m+m'\big)d\big)!}{(d!)^{m'} \big((md)!\big)^{\alpha-1}}
+ m' \frac{\big(\big((\alpha-1)m+m'\big)d\big)!}{(d!)^{m'} \big((md)!\big)^{\alpha-1}}
\end{array}}
\\
&{\normalsize
\begin{array}{l}\displaystyle
= (\alpha+m') \frac{\big(\big((\alpha-1)m+m'\big)d\big)!}{(d!)^{m'} \big((md)!\big)^{\alpha-1}}.
\end{array}}
\end{align*}
\stopcompact{footnotesize}

We claim that for any given ordered partition $Q$, if
\[
\sum_{i \in \rr} x_i \geq 1 \quad \forall\;\; \rr\in \RRR_Q,
\]
then \m{$\sum_{i=1}^{n} x_i \geq\frac{n}{r}$}.
To see this, consider the partition of $\nset$ formed by sets $U$ and $V$, where
\[
U \triangleq \bigcup_{j=1}^{m'} \uu_j,\qquad
V \triangleq \bigcup_{j=1}^{\alpha} \vv_j.
\]
Correspondingly, we partition $\RRR_Q$ into two collections of \m{$r$-subsets} $\RRR_Q^U$ and $\RRR_Q^V$, where
\[
\RRR_Q^U \triangleq \{\rr_1,\ldots,\rr_{m'}\},\qquad
\RRR_Q^V \triangleq \{\rr_{m'+1},\ldots,\rr_{m'+\alpha}\}.
\]
Observe that each element \m{$i\in U$} appears in exactly one $\uu_j$, which in turn appears in exactly $m$ of the $m'$ \m{$r$-subsets} of $\RRR_Q^U$ (namely
\m{$\rr_j,\rr_{j-1},\ldots,\rr_{j-(m-1)}$},
where
\m{$\rr_\ell \triangleq \rr_{\ell+m'}$} if \m{$\ell<1$}),
i.e.,
\[
\sum_{\rr\in\RRR_Q^U} \II{i\in\rr} = m \quad \forall\;\; i \in U.
\]
Applying Lemma~\ref{thm:lemma:RandFixedSizeSubsetCover} with \m{$S=U$}, \m{$c=m'$}, and \m{$b=m$} therefore produces
\m{$\sum_{i\in U} x_i \geq \frac{m'}{m} = \frac{r'}{r}$}.
Likewise, observe that each element \m{$i\in V$} appears in exactly one of the $\alpha$ \m{$r$-subsets} of $\RRR_Q^V$, i.e.,
\[
\sum_{\rr\in\RRR_Q^V} \II{i\in\rr} = 1 \quad \forall\;\; i \in V.
\]
Applying Lemma~\ref{thm:lemma:RandFixedSizeSubsetCover} with \m{$S=V$}, \m{$c=\alpha$}, and \m{$b=1$} therefore produces
\m{$\sum_{i\in V} x_i \geq \alpha$}.
Combining the sums of $U$ and $V$ yields
\[
\sum_{i=1}^{n} x_i
= \sum_{i\in U} x_i + \sum_{i\in V} x_i
\geq \frac{r'}{r} + \alpha
= \frac{n}{r}.
\]

Let $\RRR$ be the collection of all $\binom{n}{r}$ possible \m{$r$-subsets} of $\nset$.
As demonstrated in the proof of Theorem~\ref{thm:theorem:RandFixedSizeSubsetIffDiv}, if fewer than \m{$\frac{A}{B}$} \m{$r$-subsets} are removed from $\RRR$, then at least one collection $\RRR_Q$ can be found among the remaining \m{$r$-subsets}.
In this case, we have
\[
\frac{A}{B}
= \frac{1}{\alpha+m'} \frac{\big((\alpha m+m')d\big)!}{\big(\big((\alpha-1)m+m'\big)d\big)! (md)!}
= \frac{1}{\alpha+m'} \binom{n}{r}.
\]
Thus, some collection $\RRR_Q$ can be found among \emph{any} collection of more than
\m{$\left(1-\frac{1}{\alpha+m'}\right) \binom{n}{r}$}
distinct \m{$r$-subsets}.

We have therefore shown that if
\m{$\Ps>1-\frac{1}{\alpha+m'}$},
then any feasible allocation must satisfy
\m{$\sum_{i=1}^{n} x_i\geq\frac{n}{r}$}.
Now, $\xxr$ is a feasible allocation since it has a recovery probability of exactly $1$;
because it uses the minimum possible total amount of storage $\frac{n}{r}$, this allocation is also optimal.

Applying the preceding argument to the degenerate case of \m{$\alpha=0$} produces
\m{$\frac{A}{B} = \frac{1}{m'}\binom{n}{r}$},
which is consistent with the above expression.
\end{IEEEproof}

\begin{IEEEproof}[Proof of Corollary~\ref{thm:corollary:RandFixedSizeSubsetIffDiffDiv}]
Suppose that $n$ is a multiple of \m{$(n-r)$};
let integer \m{$\beta\geq 2$} be defined such that
\m{$n=\beta(n-r)$} $\Longleftrightarrow$
\m{$n=\frac{\beta}{\beta-1}r$}.

If \m{$\beta=2$}, then \m{$n=2r$}, i.e., $n$ is a multiple of $r$.
According to Theorem~\ref{thm:theorem:RandFixedSizeSubsetIffDiv}, $\xxr$ is an optimal allocation if and only if
\[
\Ps>1-\frac{r}{n}
=1-\frac{r}{2r}
=\frac{1}{2}
=\frac{r}{n},
\]
as required.

If \m{$\beta\geq 3$}, then $n$ is not a multiple of $r$.
We can write
\m{$n=\alpha\,r+r'$}, where
\m{$\alpha=0$} and \m{$r'=n\in\{r+1,\ldots,2r-1\}$}.
According to Theorem~\ref{thm:theorem:RandFixedSizeSubsetNotDiv}, $\xxr$ is an optimal allocation if
\[
\Ps > 1-\frac{\gcd(r,r')}{\alpha\gcd(r,r')+r'} \!
= 1-\frac{\gcd(r,n)}{n}
= 1-\frac{n-r}{n}
= \frac{r}{n}.
\]
To show that \m{$\Ps>\frac{r}{n}$} is also a necessary condition for the optimality of $\xxr$, we demonstrate that this allocation is suboptimal for any
\m{$\Ps\leq\frac{r}{n}$}.
The allocation
\m{$(1,0,\ldots,0)$}
has a recovery probability of
\m{$\binom{n-1}{r-1}\big/\binom{n}{r}$} \m{$=\frac{r}{n}$} and is therefore a feasible allocation for any
\m{$\Ps\leq\frac{r}{n}$}.
Since this allocation uses a smaller total amount of storage
\m{$1<\frac{n}{r}$}, it is a strictly better allocation than $\xxr$ for any
\m{$\Ps\leq\frac{r}{n}$}.
\end{IEEEproof}

\begin{IEEEproof}[Proof of Lemma~\ref{thm:lemma:ProbAllocUpperBoundOtherL}]
At \m{$T=\frac{n}{r}$}, the recovery probability corresponding to a particular choice of \m{$\ell\in\{1,2,\ldots,r-1\}$} is given by
\[
\Ps\left(n,r,T{=}\frac{n}{r},\ell\right)
= \PP{\BB{r}{\frac{\ell}{r}}\geq\ell}.
\]
We will prove that the above expression is at most $\frac{3}{4}$ for any \m{$\ell\in\{1,2,\ldots,r-1\}$} and \m{$r\geq 2$}
by showing that
\[
\PP{\BB{a+b}{\frac{a}{a+b}}\geq a}\leq\frac{3}{4}
\]
for any positive integers $a$ and $b$.
To do this, we consider the following three exhaustive cases separately:

\textit{Case~1}:
Suppose that \m{$a\geq 18$} and \m{$b\geq 3$}.
We will first derive an upper bound for
\m{$\PP{\BB{a+b}{\frac{a}{a+b}}\geq a}$}
by finding separate bounds for
\m{$\PP{\BB{a+b}{\frac{a}{a+b}}=a}$} and
\m{$\PP{\BB{a+b}{\frac{a}{a+b}}\geq a+1}$};
we then proceed to show that this upper bound is smaller than $\frac{3}{4}$ for any
\m{$a\geq 18$} and \m{$b\geq 3$}.

For any positive integers $a$ and $b$, we have
\begin{align}
\PP{\BB{a+b}{\frac{a}{a+b}}=a}
&= \binom{a+b}{a} \left(\frac{a}{a+b}\right)^{\!a} \left(\frac{b}{a+b}\right)^{\!b} \notag
\\
&< \frac{{\ee}^{\frac{1}{12(a+b)}}}{\sqrt{2\pi}} \sqrt{\frac{a+b}{ab}}. \label{eq:BinomialBound}
\end{align}
Inequality~\eqref{eq:BinomialBound} follows from the application of the following bound for the binomial coefficient:
\[
\binom{a+b}{a}
< \frac{\ee^{\frac{1}{12(a+b)}}}{\sqrt{2\pi}} \frac{(a+b)^{a+b+\frac{1}{2}}}{{a}^{a+\frac{1}{2}}\,{b}^{b+\frac{1}{2}}},
\]
which is derived from the following Stirling-based bounds for the factorial (see, e.g., \cite{dl:feller68probability}):
\[
\sqrt{2\pi k} \left(\frac{k}{\ee}\right)^{k}
< k!
< \sqrt{2\pi k} \left(\frac{k}{\ee}\right)^{k} \ee^\frac{1}{12k},
\qquad
k\geq 1.
\]

For any positive integers $a$ and $b$, we have
\begin{align}
\PP{\BB{a+b}{\frac{a}{a+b}}\geq a+1}\leq\frac{1}{2}, \label{eq:MedianBound}
\end{align}
which follows from the definition of the median:
The mean of the binomial random variable
\m{$\BB{a+b}{\frac{a}{a+b}}$} is \m{$(a+b)\cdot\frac{a}{a+b}$} $=a$;
since the mean is an integer, the median coincides with the mean \cite{dl:kaas80mean}.
Therefore, according to the definition of the median, we have
\[
\PP{\BB{a+b}{\frac{a}{a+b}}\leq a}\geq\frac{1}{2},
\]
which leads to inequality~\eqref{eq:MedianBound}.

Combining bounds~\eqref{eq:BinomialBound} and \eqref{eq:MedianBound} produces
\[
\PP{\BB{a+b}{\frac{a}{a+b}}\geq a}
< \frac{{\ee}^{\frac{1}{12(a+b)}}}{\sqrt{2\pi}} \sqrt{\frac{a+b}{ab}} + \frac{1}{2}
\triangleq f(a,b)
\]
for any positive integers $a$ and $b$.
Now, the upper bound \m{$f(a,b)$} is a decreasing function of both $a$ and $b$ since \m{$f(a,b)$} is a symmetric function and the partial derivative
\[
\frac{\partial}{\partial a} f(a,b)
= -\frac{6b^2+6ab+a}{12a(a+b)^2}\,\frac{{\ee}^{\frac{1}{12(a+b)}}}{\sqrt{2\pi}}\sqrt{\frac{a+b}{ab}}
\]
is negative for any \m{$a\geq 1$} and \m{$b\geq 1$}.
Thus, for any \m{$a\geq 18$} and \m{$b\geq 3$}, we have
\[
f(a,b)
\leq f(a{=}18,b{=}3)
= \frac{{\ee}^{\frac{1}{252}}}{6}\sqrt{\frac{7}{\pi}}+\frac{1}{2}
\approx 0.749773 < \frac{3}{4},
\]
which implies that
\m{$\PP{\BB{a+b}{\frac{a}{a+b}}\geq a}<\frac{3}{4}$}
for any positive integers \m{$a\geq 18$} and \m{$b\geq 3$}.

\textit{Case~2}:
Suppose that \m{$b\in\{1,2\}$}.
We will show that
\[
\textstyle
\PP{\BB{a+1}{\frac{a}{a+1}}{\geq}a}\leq\frac{3}{4}
\text{ and }
\PP{\BB{a+2}{\frac{a}{a+2}}{\geq}a}<\frac{3}{4}
\]
for any positive integer $a$.
The left-hand side of each inequality can be expanded and simplified to obtain the following:
\begin{align*}
\textstyle
\PP{\BB{a+1}{\frac{a}{a+1}}\geq a}
&\textstyle
= \frac{a^a (2a+1)}{(a+1)^{a+1}}
\triangleq f_1(a),
\\
\textstyle
\PP{\BB{a+2}{\frac{a}{a+2}}\geq a}
&\textstyle
= \frac{a^a (5a^2+10a+4)}{(a+2)^{a+2}}
\triangleq f_2(a).
\end{align*}
The first derivatives of \m{$f_1(a)$} and \m{$f_2(a)$}, which are given by
\begin{align*}
\textstyle
f'_1(a)
&\textstyle = \frac{a^a}{(a+1)^{a+1}} \left\{2-(2a+1)\ln\left(\frac{a+1}{a}\right)\right\},
\\
\textstyle
f'_2(a)
&\textstyle
 = \frac{a^a}{(a+2)^{a+2}} \left\{(10a+10)-(5a^2+10a+4)\ln\left(\frac{a+2}{a}\right)\right\},
\end{align*}
can be shown to be negative for any \m{$a\geq 1$}.
Since
\m{$f_1(a{=}1)=\frac{3}{4}$},
\m{$f_2(a{=}1)=\frac{19}{27}<\frac{3}{4}$},
and both \m{$f_1(a)$} and \m{$f_2(a)$} are decreasing functions of $a$ for any \m{$a\geq 1$}, it follows that
\m{$f_1(a)\leq\frac{3}{4}$} and \m{$f_2(a)<\frac{3}{4}$}
for any positive integer $a$, as required.

\textit{Case~3}:
Suppose that \m{$a\in\{1,2,\ldots,17\}$}.
We will describe our approach for \m{$a=1$} and \m{$a=2$};
the proofs for the other 15 cases are similar, and can be verified with the help of a computer.
We will show that
\[
\textstyle
\PP{\BB{b+1}{\frac{1}{b+1}}{\geq}1}\leq\frac{3}{4}
\text{ and }
\PP{\BB{b+2}{\frac{2}{b+2}}{\geq}2}<\frac{3}{4}
\]
for any positive integer $b$.
The left-hand side of each inequality can be expanded and simplified to obtain the following:
\begin{align*}
\textstyle
\PP{\BB{b+1}{\frac{1}{b+1}}\geq 1}
&\textstyle
= 1-\frac{b^{b+1}}{(b+1)^{b+1}}
\triangleq g_1(b),
\\
\textstyle
\PP{\BB{b+2}{\frac{2}{b+2}}\geq 2}
&\textstyle
= 1-\frac{b^{b+1} (3b+4)}{(b+2)^{b+2}}
\triangleq g_2(b).
\end{align*}
The first derivatives of \m{$g_1(b)$} and \m{$g_2(b)$}, which are given by
\begin{align*}
\textstyle
g'_1(b)
&\textstyle = \frac{b^b}{(b+1)^{b+1}} \left\{b\ln\left(\frac{b+1}{b}\right)-1\right\},
\\
\textstyle
g'_2(b)
&\textstyle = \frac{b^b}{(b+2)^{b+2}} \left\{(3b^2+4b)\ln\left(\frac{b+2}{b}\right)-(6b+4)\right\},
\end{align*}
can be shown to be negative for any \m{$b\geq 1$}.
Since
\m{$g_1(b{=}1)=\frac{3}{4}$},
\m{$g_2(b{=}1)=\frac{20}{27}<\frac{3}{4}$},
and both \m{$g_1(b)$} and \m{$g_2(b)$} are decreasing functions of $b$ for any \m{$b\geq 1$}, it follows that
\m{$g_1(b)\leq\frac{3}{4}$} and \m{$g_2(b)<\frac{3}{4}$}
for any positive integer $b$, as required.
\end{IEEEproof}

\begin{IEEEproof}[Proof of Theorem~\ref{thm:theorem:ProbAllocMaxSpreadOptT}]
We have already established that the choice of \m{$\ell=r$} is optimal for any \m{$T\geq\frac{n}{r}$};
it therefore suffices to show that \m{$\ell=r$} is also optimal for any
\m{$T\in\left[\frac{n}{r}\left(\frac{3}{4}\right)^{\frac{1}{r}},\frac{n}{r}\right)$}.

The recovery probability corresponding to any
\m{$\ell\in\{1,2,\ldots,r\}$} is given by
\[
\textstyle
\Ps(n,r,T,\ell)
= \PP{\BB{r\vphantom{\Big|}}{\min\!\left(\frac{\ell T}{n},1\right)}\geq\ell},
\]
which is a nondecreasing function of $T$ since
\m{$\min\!\left(\frac{\ell T}{n},1\right)$}
either increases or remains constant at $1$ as $T$ increases.
More precisely, \m{$\Ps(n,r,T,\ell)$} is an increasing function of $T$ on the interval \m{$\left(0,\frac{n}{\ell}\right)$};
for higher values of $T$, the function saturates at $1$.
We can verify this claim by checking that the partial derivative
\[
\frac{\partial}{\partial p} \PP{\BB{r}{p}\geq\ell}
= \ell\binom{r}{\ell} p^{\ell-1} (1-p)^{r-\ell}
\]
is positive for any \m{$p\in(0,1)$}.

Now, the recovery probability corresponding to the choice of \m{$\ell=r$} at \m{$T=\frac{n}{r}\left(\frac{3}{4}\right)^{\frac{1}{r}}$} is given by
\[
\textstyle
\Ps\left(n,r,T{=}\frac{n}{r}\left(\frac{3}{4}\right)^{\frac{1}{r}},\ell{=}r\right)
= \PP{\BB{r}{\left(\frac{3}{4}\right)^{\frac{1}{r}}}\geq r}
= \displaystyle\frac{3}{4}.
\]
Since \m{$\Ps(n,r,T,\ell)$} is a nondecreasing function of $T$, we have
\begin{align*}
\Ps(n,r,T,\ell{=}r) &\geq\frac{3}{4}
\quad \text{for any } T\geq\frac{n}{r}\left(\frac{3}{4}\right)^{\frac{1}{r}}.
\intertext{On the other hand, for any \m{$\ell\in\{1,2,\ldots,r-1\}$}, we have}
\Ps(n,r,T,\ell) &\leq\frac{3}{4}
\quad \text{for any }
T\leq\frac{n}{r}, \notag
\end{align*}
from the upper bound of Lemma~\ref{thm:lemma:ProbAllocUpperBoundOtherL}.
It therefore follows that the choice of \m{$\ell=r$} is optimal for any
\m{$T\in\left[\frac{n}{r}\left(\frac{3}{4}\right)^{\frac{1}{r}},\frac{n}{r}\right)$},
as required.
\end{IEEEproof}

\begin{IEEEproof}[Proof of Corollary~\ref{thm:corollary:ProbAllocMaxSpreadOptPs}]
Theorem~\ref{thm:theorem:ProbAllocMaxSpreadOptT} already demonstrates that the choice of \m{$\ell=r$} is optimal for any
\m{$T\geq\frac{n}{r}\left(\frac{3}{4}\right)^{\frac{1}{r}}$};
we will proceed to show that a recovery probability of at least $\frac{3}{4}$ is \emph{not} achievable for any
\m{$T<\frac{n}{r}\left(\frac{3}{4}\right)^{\frac{1}{r}}$}.

Recall from the proof of Theorem~\ref{thm:theorem:ProbAllocMaxSpreadOptT} that the recovery probability \m{$\Ps(n,r,T,\ell)$} corresponding to any \m{$\ell\in\{1,2,\ldots,r\}$} is an increasing function of $T$ on the interval \m{$\left(0,\frac{n}{\ell}\right)$}.
Thus, for the choice of \m{$\ell=r$}, the function \m{$\Ps(n,r,T,\ell{=}r)$} increases wrt $T$ on the subinterval
\m{$\left(0,\frac{n}{r}\left(\frac{3}{4}\right)^{\frac{1}{r}}\right]\subset\left(0,\frac{n}{r}\right)$};
since \m{$\Ps\left(n,r,T{=}\frac{n}{r}\left(\frac{3}{4}\right)^{\frac{1}{r}},\ell{=}r\right)=\frac{3}{4}$},
it follows that
\begin{align*}
\Ps(n,r,T,\ell{=}r) &<\frac{3}{4}
\quad \text{for any }
T<\frac{n}{r}\left(\frac{3}{4}\right)^{\frac{1}{r}}.
\intertext{On the other hand, for any
\m{$\ell\in\{1,2,\ldots,r-1\}$}, the function \m{$\Ps(n,r,T,\ell)$} increases wrt $T$ on the subinterval
\m{$\left(0,\frac{n}{r}\right]\subset\left(0,\frac{n}{\ell}\right)$};
since \m{$\Ps\left(n,r,T{=}\frac{n}{r},\ell\right)\leq\frac{3}{4}$}
according to Lemma~\ref{thm:lemma:ProbAllocUpperBoundOtherL}, it follows that}
\Ps(n,r,T,\ell) &<\frac{3}{4}
\quad \text{for any }
T<\frac{n}{r}.
\end{align*}
Hence, the optimal recovery probability for any
\m{$T<\frac{n}{r}\left(\frac{3}{4}\right)^{\frac{1}{r}}$}
is strictly less than $\frac{3}{4}$.
\end{IEEEproof}

\section*{Acknowledgment}

The authors would like to thank \m{Brighten} \m{Godfrey} and \m{Robert} \m{Kleinberg} for introducing the problem to them and for sharing their insights.
The authors also thank \m{Dimitris} \m{Achlioptas} for the interesting discussions.

\ifCLASSOPTIONcaptionsoff
  \newpage
\fi

\IEEEtriggeratref{30}

\end{document}